\documentclass[twocolumn]{aastex701}

\usepackage{amsmath}
\usepackage{bm}
\usepackage{siunitx}
\usepackage{cleveref}
\usepackage{caption}

\newcommand{\lcdm}{$\Lambda$CDM}

\begin{document}

\title{Testing \lcdm\ versus dynamical dark energy in one year: \\ A DESI spectroscopic follow-up program for Rubin supernovae}

\newcommand{\berkeley}{\affiliation{Department of Physics, University of California, Berkeley, CA 94720, USA}}
\newcommand{\lbnl}{\affiliation{E.O. Lawrence Berkeley National Laboratory, 1 Cyclotron Rd., Berkeley, CA, 94720, USA}}

\author[orcid=0009-0004-8106-9452]{Jannik Truong}
\berkeley
\email[show]{jannik@berkeley.edu}

\author[]{Greg Aldering} 
\lbnl
\email{galdering@lbl.gov}

\author[orcid=0000-0002-4436-4661]{Saul Perlmutter}
\berkeley\lbnl
\email{saul@lbl.gov}

\author[orcid=0000-0001-5402-4647]{David Rubin}
\affiliation{Department of Physics and Astronomy, University of Hawai`i at M{\=a}noa, Honolulu, Hawai`i 96822}
\email{drubin@hawaii.edu}

\author[orcid=0000-0002-5042-5088]{David Schlegel}
\lbnl
\email{djschlegel@lbl.gov}

\begin{abstract}

Combined cosmological probes currently indicate that best-fit values in the $w_0-w_a$ parametrization of dynamical dark energy deviate from \lcdm\ by $\sim3\sigma$. In this work, we present a supernova survey capable of measuring dynamical dark energy at the $>5\sigma$ level with just one year of data, starting in 2027.
We first show that with the present values of $w_0$ and $w_a$, new SNe~Ia at redshifts $z\lesssim0.6$ near dark energy-matter equality would add the most constraining power. This is well within reach of the Vera C. Rubin Observatory and the Dark Energy Spectroscopic Instrument (DESI).
Because cosmology measurements with SNe~Ia quickly become systematics-limited, we focus on eliminating key systematics by using only a spectroscopically confirmed and volume-limited sample. 
In our proposed survey, SN alerts from Rubin would actively re-prioritize the scheduling of already-planned DESI tile visits. This would yield 7\,500 near-peak transient spectra in one year without delaying DESI's primary survey. 
We forecast that if current best-fit $w_0-w_a$ values persist, combining just our volume-limited subset of 2\,300 new SNe~Ia at $z<0.3$ with current SN, BAO, and CMB data would push the tension with \lcdm~beyond $5\sigma$. This applies across a wide range of assumed uncertainties.
To further circumvent systematics, we explore how DESI enables spectroscopic standardization via machine learning, offering a path toward a cosmology measurement independent of light-curve-based standardization. Finally, we discuss how early results from this program could inform future dark energy experiments.

\end{abstract}



\section{Introduction} \label{sec:intro}

\begin{figure}[]
    \centering
    \includegraphics[width=0.49\textwidth]{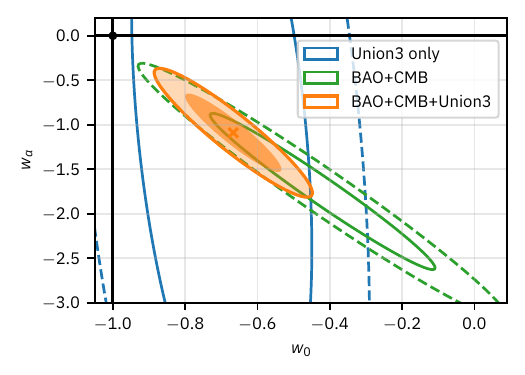}
    \caption{Reconstructed $1\sigma$ and $2\sigma$ Gaussian contours for $w_0$ and $w_a$ from different datasets: Union3 \citep{Rubin2025Union3} SNe~Ia only (blue), DESI DR2 BAO+Planck CMB without SNe (green, \citet{DESIDR2BAO2025}), and all three combined (orange). When combining all three datasets, \lcdm~($w_0=-1$ and $w_a=0$, top left black point) is $3.8\sigma$ from the best-fit values.
    \label{fig:ellipses_baseline}}
\end{figure}

Type Ia supernovae (SNe~Ia) enabled the discovery of dark energy \citep{Riess1998, Perlmutter1999}. For over two decades, a cosmological constant remained consistent with observations. But recent analyses suggest that dark energy may be dynamical. SN~Ia data were the first to show hints of an evolving equation of state \citep{Rubin2025Union3}, and Baryon Acoustic Oscillation (BAO) measurements from the Dark Energy Spectroscopic Instrument \citep{DESIDR2BAO2025} have significantly strengthened the case: Current best-fit values for the $w_0$ and $w_a$ parameters of dynamical dark energy are $\sim 3\,\sigma$ away from \lcdm~when data from SNe, BAO and the cosmic microwave background (CMB) are combined (\Cref{fig:ellipses_baseline}). 

The current level of resources being focused on the dark energy question is unparalleled: Large-scale ``Stage-IV'' \citep{Albrecht2006DETF} experiments such as DESI \citep{DESI2016}, Euclid \citep{Euclid2025} and the Vera C. Rubin Observatory \citep{Ivezic2019LSST} are taking data with the goal of measuring dark energy with more precision than ever before. More surveys will soon launch, including the Roman Space Telescope \citep{Spergel2015Roman}, Lazuli Space Telescope \citep{Roy2026Lazuli}, and potentially DESI-II. At the same time, hints of evolving dark energy motivate a rapidly expanding body of theoretical work aimed at explaining these new observations \citep{Shlivko2024,Lodha2025,Wolf2025}.

Our goal with this paper is to find a fast and timely means of testing \lcdm\ versus dynamical dark energy at the $>5\sigma$ level with supernovae --- by marshalling the tremendous resources provided by current and upcoming experiments.
Reaching $5\sigma$ confidence that dark energy is dynamical would both add momentum to theoretical efforts and steer the design of future dark energy surveys, many of which were conceptualized without considering a potential departure from the \lcdm\ paradigm.

Given that CMB and DESI BAO data alone are projected to only reach a $3.9\sigma$ confidence on dynamical dark energy even with the extension of DESI (Haruki Ebina, priv. comm.), Type Ia supernovae remain essential for crossing the $5\sigma$ discovery threshold.

\subsection{A brief review of supernova cosmology}

Supernova cosmology uses the standardizable luminosity of Type Ia supernovae to map the expansion history of the universe, thereby measuring the strength of dark energy over cosmic time.

Transients such as SNe~Ia are discovered by repeat observations of the same patch of the sky. Ideally, a discovered transient receives live spectroscopic follow-up, giving a redshift measurement and classification. SNe~Ia are uniquely identifiable by strong spectral features of singly ionized silicon, SiII, as well as the lack of hydrogen and helium lines found in many core-collapse supernovae \citep{Filippenko1997}.

Supernova light curves, obtained by difference imaging in multiple filters, are important for standardization: Unstandardized absolute magnitudes have a root-mean-square scatter of \SI{0.4}{mag}, corresponding to a distance uncertainty of $18\%$. To improve cosmological distance measurements, SNe~Ia are empirically standardized \citep{phillips1993} using parameters from fits to light curve models such as \texttt{SALT} \citep{Guy2007, Kenworthy2021} or \texttt{SNooPY} \citep{Burns2011}. This procedure extracts a peak magnitude $m_B$, a stretch parameter $x_1$, and a color parameter $c$. In the simplest approach, these parameters are combined via the Tripp relation \citep{Tripp1998}
\begin{align}
    \label{eq:tripp}
    m_B^{\mathrm{corr}} = m_B + \alpha x_1 - \beta c
\end{align}
where $\alpha$ and $\beta$ are nuisance parameters fit from the data. The stretch parameter captures the width of the light curve: SNe with broader, slower-declining light curves ($x_1 > 0$) are intrinsically more luminous and thus appear brighter than average. The color parameter captures reddening: redder SNe ($c > 0$) appear fainter, whether due to dust extinction or intrinsic variation. Both effects must be corrected for to recover the standardized magnitude, and thus distance. After standardization, the scatter in SN~Ia magnitudes typically reduces to $\sim$\SI{0.15}{mag}, or a distance uncertainty of $\sim 7\%$. This residual scatter is often called ``unexplained dispersion''.

The corrected supernova magnitudes and redshifts are fit to the theoretical prediction
\begin{align}
    m_B(z) = 5\log_{10}\mathcal{D}_L(z|w_0,w_a,\Omega_M)+\mathcal{M}
\end{align}
where $\mathcal{D}_L=H_0d_L/c$ is the dimensionless luminosity distance. Constants are absorbed into the nuisance parameter $\mathcal{M}=M+5\log_{10}(c/H_0/10\,\rm{pc})$, where $M$ is the mean absolute magnitude of the supernova (e.g. \citealt{Perlmutter1997}). In practice, fits are often performed to the distance moduli $\mu=m-M$, which is an entirely equivalent procedure. 

To evaluate dynamical dark energy, we assume a flat $w_0w_a$CDM cosmology and use the Chevallier-Polarski-Linder parameterization \citep{Chevallier2001,Linder2003}
\begin{align}
    \label{eq:w0wa}
    w(a)=w_0+w_a(1-a)
\end{align}
to describe the evolving dark energy equation of state $w=p/\rho$, where setting $w_0=-1$ and $w_a=0$ recovers \lcdm.

\subsection{Improving beyond current supernova datasets}

Three major supernova datasets exist, all containing 1500 to 2000 SNe~Ia. Pantheon+ \citep{Brout2022Pantheon} and Union3 \citep{Rubin2025Union3} compile spectroscopically confirmed SNe~Ia across various programs from the literature. There is significant overlap in the included SNe, even though analysis techniques differ. In contrast, the $z>0.1$ SNe from the Dark Energy Survey \citep{DES5Y2024, Popovic2025b} are all obtained by the same instrument, achieving more consistent photometric calibration. However, the DES sample is partially contaminated by non-Ia supernovae, as a large part of the sample was not confirmed spectroscopically by examining spectral features, but only photometrically classified based on their light curves, introducing new uncertainties.

The next step for supernova cosmology is thus to combine the strengths of these surveys, eliminating critical systematics: A photometrically homogeneous sample combined with rigorous spectroscopic follow-up would be a major improvement --- even without increasing the total number of supernovae significantly.

However, even a sample like that must contend with selection effects. In magnitude-limited surveys, Malmquist bias \citep{Malmquist1925} occurs because intrinsically brighter supernovae are preferentially detected and followed up, skewing distance measurements on the Hubble diagram. To avoid this, we aim for a well-defined volume-limited subsample within the broader survey. Because a volume-limited sample includes all supernovae within a given distance regardless of their intrinsic brightness, it natively circumvents this bias. This eliminates the need for selection-effect corrections \citep{Kessler2017BBC} that themselves introduce correlated uncertainties.

\subsection{The Rubin Observatory and the need for spectroscopic follow-up}

The Vera C. Rubin Observatory's Legacy Survey of Space and Time (LSST, \citealt{Ivezic2019LSST}), a wide-field ground-based optical survey with broad science goals, is well positioned to deliver the required photometric homogeneity described above. Beginning in 2026, Rubin will deliver unprecedented statistics, measuring an estimated hundreds of thousands of new SNe~Ia over ten years \citep{LSSTScienceBook, LSSTDESCSRD}; cosmological analyses using these SNe will thus be limited by systematic uncertainties.

\begin{figure}[]
    \centering
    \includegraphics[width=0.49\textwidth]{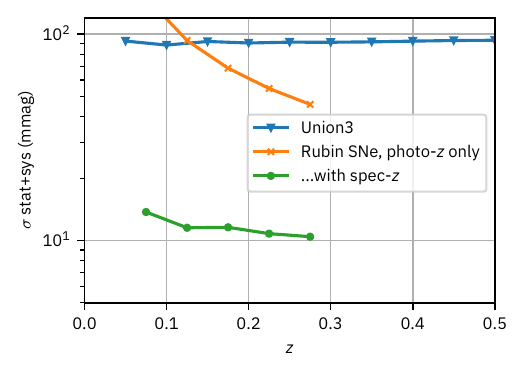}
    \caption{Comparison of uncertainties of binned distance moduli between $\sim3000$ simulated volume-limited $z<0.3$ LSST supernovae with (green) and without (orange) DESI follow-up, and Union3 (blue). We plot the square roots of the diagonal elements of the distance modulus covariance matrix, and $\sigma_{\rm corr}$ correspond to the first off-diagonal. Without spectroscopic follow-up, the new statistical power of LSST SNe can hardly be utilized as systematics approach values comparable to Union3. The photo-$z$-induced uncertainties are estimated based on a systematic $\sigma_z=0.005$.
    \label{fig:photoz_comparison}}
\end{figure}
Rubin's photometrically classified samples will be contaminated with non-Ia supernovae, which are generally fainter.
Thus, live spectroscopic follow-up is necessary to classify transients with certainty. Indeed, simulations show that photometric mis-classification alone systematically shifts the resulting cosmology \citep{Mitra2025, Malz2025}, an effect that could soon become the limiting factor: \citet{Vincenzi2023} find a bias of order 0.1 in $w_a$ from mis-classification alone --- half a standard deviation or more with the soon-expected increase in precision.\footnote{With BAO data from DESI DR2, the total uncertainty on $w_a$ already drops below 0.3 \citep{DESIDR2BAO2025}.}

Likewise, because Rubin only provides photometry measurements, SN redshifts will need to be estimated photometrically from the SN and its host galaxy. At low redshift ($z<0.5$), errors propagate\footnote{At higher redshifts $z>0.5$, (anti-)correlations between the photometric redshift and light curve parameters partially compensate for inaccurate photo-$z$ measurements \citep{Mitra2023, Kessler2025}. In this case, the final impact on $\sigma_\mu$ is thus smaller than expected by uncorrelated error propagation. This effect does not occur at low redshifts $z<0.5$, precisely where $\sigma_{\mu}^{\rm phot}$ quickly grows with $1/z$.} to the measured distance moduli as 
\begin{align}
\label{eq:photo-z}
    \sigma_{\mu}^{\rm phot}=\frac{d\mu}{dz}\,\sigma_z\approx\frac{5}{\ln10}\,\frac{\sigma_z}{z}.
\end{align}
Even with a rather optimistic systematic $\sigma_{z}^{\rm phot}=0.005$ \citep{LSSTDESCSRD}, the measured distance moduli at $z=0.1$ may be systematically off by $\sigma_{\mu}^{\rm phot}>\SI{0.1}{mag}$, single-handedly exceeding the entire systematics budget of Union3 (\Cref{fig:photoz_comparison}). Additionally, a small fraction of photometric host redshifts will be catastrophic outliers \citep{Schmidt2020photoz}. In comparison, spectroscopic redshifts from the host galaxy are accurate to $\sigma_z^{\rm spec}\approx10^{-5}$ \citep{Lan2023}.

Even for the small subset of host galaxies with a previously-known spectroscopic redshift, a live supernova redshift helps to correctly match the SN to its host.\footnote{Host galaxy redshifts are more accurate since galaxy emission lines are narrow compared to supernova absorption lines, which are blueshifted to varying degrees by the expanding ejecta. Galaxy emission lines are often visible in live supernova spectra, so live follow-up can eliminate the need for a dedicated host galaxy measurement. For example spectra, see \citet{Hall2026}.} Otherwise, incorrect host associations in photometry-only surveys lead to redshift misassignments that bias the inferred dark energy equation of state \citep{Qu2024, Mitra2025}.


When combining all the above effects, \citet{Mitra2025} forecast that in photometry-only Rubin analyses, the aggregate biases on $w_0$ and $w_a$ will each exceed a full standard deviation at the precision required to reach $5\sigma$ from \lcdm.

In short, photometry-only analyses are systematically vulnerable on both dimensions of the Hubble diagram: redshift errors on the $x$-axis and mis-classification biases on the $y$-axis. Live spectroscopy directly resolves both weaknesses, and would thus significantly strengthen a potential $5\sigma$ discovery claim.

\subsection{Cross-survey synergies and a window of opportunity for DESI}

The current largest-scale plan to follow up Rubin transients is the Time Domain Extragalactic Survey (TiDES, \citealt{Frohmaier2025TiDES}) of the 4-metre Multi-Object Spectroscopic Telescope (4MOST, \citealt{deJong20194MOST}). Over five years, it will provide around 10\,000 spectra of SNe~Ia, and over 100\,000 accurate host galaxy redshifts. Yet, this program is not optimized for supernova cosmology: Telescope pointings are scheduled without consideration of live transients, so they are only observed if they happen to fall into a pre-planned pointing. SNe~Ia can thus not be observed near peak brightness consistently. Fixed exposure times that do not scale according to observing conditions also lead to inconsistent depths. Both effects exacerbate spectroscopic Malmquist bias, where intrinsically brighter SNe are more likely to pass signal-to-noise cuts, and are thus preferentially classified. \citet{Frohmaier2025TiDES} show that this effect occurs for $z > 0.15$ for TiDES.

Recent work has called for greater coordination among dark energy experiments in general \citep{Leauthaud2025}. Efforts to identify survey synergies for supernova cosmology have been explored for other instrument combinations \citep{Rose2021a}, but not for DESI and Rubin specifically. DESI \citep{DESI2016} is currently the most powerful ground-based multi-fiber spectrograph, capable of taking up to 5000 spectra in a single exposure, making it a prime candidate for a systematic follow-up program of Rubin transients. Furthermore, DESI achieves consistent depths by scaling an effective exposure time to real observing conditions \citep{Schlafly2023}. This dynamic exposure strategy mitigates spectroscopic Malmquist bias and thus makes DESI uniquely capable of building the rigorous, volume-limited subsample required for precision cosmology.

Coordinating these two experiments would make better use of available resources to extract more information from SNe --- ideally through a little-to-no-cost integration into DESI's existing survey --- and thereby contribute towards their shared main science goal of measuring dark energy.

A critical window of opportunity exists in 2027, when DESI overlaps with Year 2 of Rubin, which is when the rolling cadence begins in Rubin's Wide Fast Deep (WFD) survey, producing well-sampled SN light curves. At the same time, DESI already began extending its sky coverage into the southern hemisphere where Rubin operates. If the Rubin WFD starts rolling by April 2027, the overlap lasts for 1.75 years: In December 2028, the presently-funded phase of DESI will end, making timely coordination essential for maximizing scientific return.

\subsection{Paper outline}

The outline of the paper is as follows:
Section~\ref{sec:toymodel} employs a toy model to demonstrate that in the current data landscape, the greatest leverage on dynamical dark energy comes from new SNe~Ia at redshifts below $z\approx 0.6$. Section~\ref{sec:sims} describes our emulation of Rubin's Wide Fast Deep (WFD) and details our simulation framework as well as the treatment of systematic uncertainties. Section~\ref{sec:DESI} presents the design of our DESI spectroscopic follow-up program, including simulations for the expected depths for standard exposures, scheduling, and expected yields. Section~\ref{sec:Results} reports our cosmological forecasts and explores the impact of systematics. Section~\ref{sec:outlook} discusses possible future directions. We conclude in Section~\ref{sec:Conclusion}.
\section{Not all redshifts are created equal: where new supernovae help most}\label{sec:toymodel}

\begin{figure*}[t]
    \centering
    \includegraphics[width=0.99\textwidth]{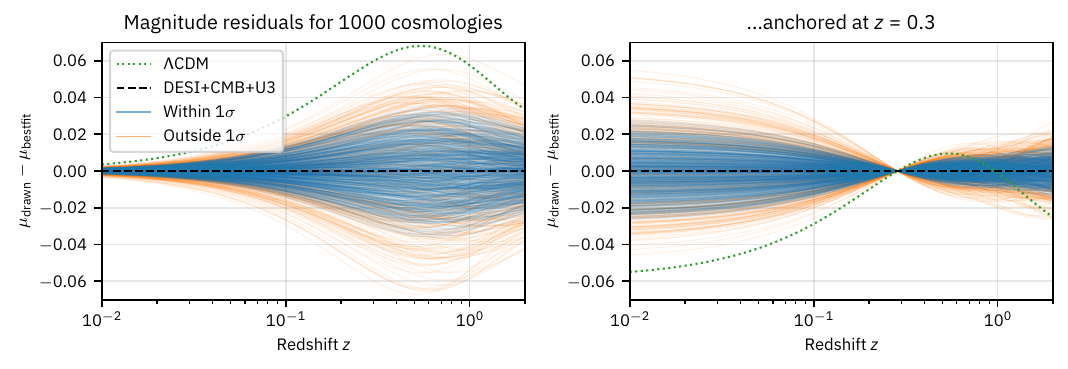}
    \caption{We sample different cosmologies according to the current (DESI DR2+Planck+Union3) best-fitting probability distributions for $w_0, w_a, \Omega_m$, compute the theoretical distance moduli and compare them against the best-fit cosmology (black dashed line). On the left panel, we can see how measuring additional SNe~Ia at redshifts $z\approx0.55$ would currently be the most helpful for constraining dynamical dark energy. On the right panel, residuals are anchored at $z=0.3$, for example, by an additional high-precision measurement. Now, low-$z$ SNe become the next best place to measure. So these plots highlight not only that there is an optimal redshift to measure at, but that a stronger medium-redshift anchor is required to make additional low-redshift supernovae more informative.
    \label{fig:hubble_plot}}
\end{figure*}

\begin{figure*}[]
    \centering
    \includegraphics[width=0.99\textwidth]{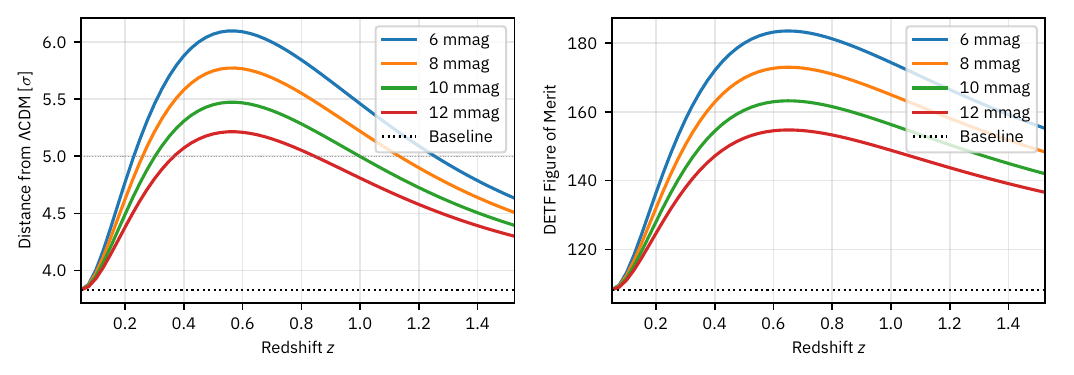}
    \caption{Distance from \lcdm\ given by the number of standard deviations, and Figure of Merit when adding supernovae at one redshift bin $z$ with different total uncertainties to DESI DR2+Planck+Union3 and low-$z$ SNe. Total uncertainties include statistical and systematic errors, although statistical errors typically become subdominant for $10^3$ to $10^4$ SNe~Ia. We find that measuring around $z\approx0.6$ increases the distance from \lcdm~and the Figure of Merit most.
    \label{fig:z_1d_grid+lowz}}
\end{figure*}

\begin{figure}[]
    \centering
    \includegraphics[width=0.49\textwidth]{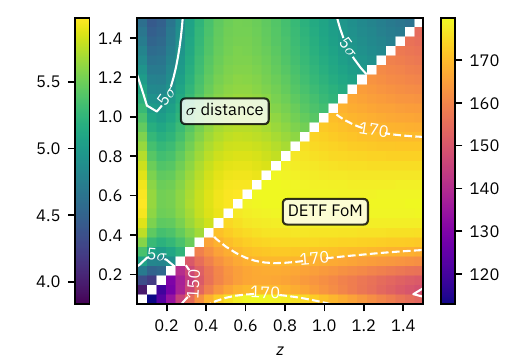}
    \caption{Distance from \lcdm~and Figure of Merit when adding supernovae at two redshift bins ($z_1$ and $z_2$) with $10\rm mmag$ total uncertainty each to DESI DR2+Planck+Union3 and low-$z$ SNe. Two measurements at medium redshifts $z\approx0.6$ and the combination of low-$z$ and medium-$z$ are considered optimal.
    \label{fig:z_grid_current+lowz}}
\end{figure}

The constraining power of new supernovae on dark energy depends on their redshifts. Intuitively, one might think that high-redshift supernovae are the most beneficial: The luminosity distance is an integral over redshift, so going to higher redshifts would seem to include a larger chunk of the expansion history. 

But when comparing distance moduli from different $w_0w_a\rm CDM$ cosmologies drawn from current best-fit probabilities (\Cref{fig:hubble_plot}), we similarly find the biggest variation at medium redshifts --- $z\approx0.55$, implying that the SNe necessary for testing \lcdm\ versus $w_0w_a\rm CDM$ are well within Rubin and DESI capabilities.

We briefly investigate the following: Given the current data from BAO, CMB and SNe, which redshifts should one ideally target with a new supernova survey intended to test constant versus dynamical dark energy? Which redshifts reduce the size of the $w_0-w_a$ error ellipse --- quantified by the Dark Energy Task Force Figure of Merit (DETF FoM, \citealt{Albrecht2006DETF}, and Appendix \ref{sec:fisher}) --- the most, or, alternatively, offer the largest increase in the distance from \lcdm?

\subsection{Toy model}

In our toy model, we explore the addition of supernova measurements concentrated into a single redshift.\footnote{This naturally leads to a smaller constraining power compared to multiple bins that can ``pinch'' the Hubble diagram at multiple redshifts.} We assume that this binned measurement has a combined baseline uncertainty of \SI{10}{mmag}. This allows for a conservative accounting of statistical and systematic uncertainties. For example, the statistical uncertainty would be \SI{5}{mmag} for 1000 SNe with an unexplained dispersion of \SI{0.15}{mag}. Modern calibration of filter zero-points traces back to the CALSPEC system of fundamental white dwarf stars \citep{Bohlin2014}. This is typically accurate up to \SI{5}{mmag} (Union3, \citealt{Rubin2025Union3}) or less (DES-Dovekie, \citealt{Popovic2025a}). And the simulated covariances in \citet{Rubin2025} reach values below \SI{10}{mmag} consistently out to $z\approx1.5$, even with comprehensive Union3-like systematics.

We compute the corresponding Fisher matrix \citep{Fisher1922}
\begin{align}
    F_{ab}^{\rm SN} = \sum_{i,j} \frac{\partial \mu(z_i)}{\partial p_a}\Bigr|_{\rm fid}\,\mathcal{C}^{-1}(z_i,z_j)\,\frac{\partial \mu(z_j)}{\partial p_b}\Bigr|_{\rm fid}
    \label{eq:FisherSN}
\end{align}
where $\bm{p}=\{w_0,w_a,\Omega_M,\mathcal{M}\}$ are the cosmological parameters, $\mathcal{C}$ is the distance modulus covariance matrix ($1\times1$ for a single binned measurement) with units $\rm mag^2$, and add that to current data as described in the following.

\subsection{Baseline data and choice of fiducial cosmology}

The baseline dataset assumed throughout this paper is a combination of BAO, CMB and SNe. For the supernovae, we use the Union3 dataset \citep{Rubin2025Union3}, as its systematics treatment is most similar to the simulation we will use in this paper. We also include a low-$z$ supernova sample, for example from nearby supernova surveys such as ZTF \citep{Bellm2019ZTF, Ginolin2025}, LS4 \citep{Miller2025LS4}, or a potential Rubin short-exposure micro survey (\S\ref{sec:outlook}).

We calculate a Fisher matrix using the parameter covariance matrices from DESI DR2 \citep{DESIDR2BAO2025}\footnote{\url{https://data.desi.lbl.gov/public/papers/y3/bao-cosmo-params/cobaya/base_w_wa/}}, and add to that a Fisher matrix (\Cref{eq:FisherSN}) from a \SI{10}{mmag} measurement uncertainty at $z=0.05$ as a conservative estimate of a low-$z$ SN sample. We err on the conservative side deliberately so that our results do not presuppose a perfectly calibrated low-redshift sample. A well-calibrated sample --- especially if on the same photometric system as the $z>0.1$ Rubin data --- would have more cosmological impact than this single binned \SI{10}{mmag} measurement. Some technical details of the Fisher matrix calculations are explained in Appendix \ref{sec:fisher}.

Typically, surveys evaluate their dynamical dark energy constraining power by computing the Figure of Merit at a fiducial \lcdm\ cosmology ($w_0=-1, w_a=0$). But since we are interested in probing the hints of a potentially time-evolving dark energy \citep{Rubin2025Union3, DES5Y2024, DESIDR2BAO2025}, we evaluate the FoM at the DESI+CMB+Union3 best-fit values of $w_0=-0.667, w_a=-1.09, \Omega_m=0.327$ \citep{DESIDR2BAO2025}.\footnote{This leads to typically $10\%$ lower FoM values than when evaluated at \lcdm\ for the same distance modulus covariances \citep{Rubin2025}.} And now we can probe the discovery potential of dynamical dark energy directly by using the projected distance from \lcdm\ as an additional metric.

The current baseline values for DESI+CMB+Union3 and low-$z$ SNe are $\rm{FoM}=108$ and $3.8\sigma$ from \lcdm. These values are within 1\% with or without the addition of new low-$z$ SNe, highlighting the fact that new mid-$z$ SNe are required to make a low-$z$ sample cosmologically useful (\Cref{fig:hubble_plot}).

\subsection{Results}

We plot the resulting FoM values and distances from \lcdm\ as a function of redshift at which a single measurement is added in \Cref{fig:z_1d_grid+lowz}. Adding SNe at redshifts 0.5 to 0.6 leads to the largest gains in FoM and \lcdm-distance. Furthermore, observing costs increase sharply towards higher redshifts.\footnote{To maintain a constant signal-to-noise ratio for faint supernovae, where noise is dominated by the background, the required exposure time $t$ scales roughly as $t\propto d_L^4$.} Thus, an optimized, cost-conscious survey must have a lower ``optimal'' redshift.

The ``optimal'' redshift can be understood as the interplay of two effects: Above $z\approx0.3$, the universe becomes matter-dominated, so dark energy dynamics impact observables such as supernova luminosity distances less. However, when constraining a potentially time-evolving dark energy, it is desirable to go back in time far enough to establish a long lever arm on the Hubble diagram. This naturally pushes the maximally informative redshift out to $z\approx0.6$, irrespective of how the dark energy equation of state is parametrized.\footnote{Additionally, \citet{Lodha2025} points out that the CPL parametrization $(w_0, w_a)$ is flexible enough to explain DESI BAO results, and that the preference for evolving dark energy is not driven by this parametrization.} This is consistent with \citet{Spergel2002}, who found similar optimal redshifts for dark energy with a constant equation of state $w$. 

We also plot the FoM values and \lcdm-distances as a function of two redshifts, each containing an independent measurement, in \Cref{fig:z_grid_current+lowz}. The results are mostly the same --- going to high redshifts is not better when optimizing for FoM, and is even disfavored when optimizing for distance from \lcdm. Redshifts around $z\approx0.6$ are preferred. And an additional optimal combination arises: adding more low-redshift supernovae at $z=0.05$, combined with supernovae at medium redshifts $z\approx 0.6$, is also optimal. Thus, there is merit in having multiple high-quality low-$z$ supernova samples, as these are especially powerful at constraining late-time dark energy $w_0$.

When adding a third independent measurement, no other optimal redshift combination emerges.

The key takeaway is that new high-precision low- to medium-redshift ($z<0.6$) supernovae are both necessary and sufficient for testing \lcdm\ versus dynamical dark energy at a $5\sigma$ level if the best-fit values do not change significantly. They will provide the most constraining power and are cost-effective to spectroscopically follow up.
\section{Rubin light curve simulation} \label{sec:sims}

Rubin/LSST's Wide Fast Deep (WFD) will use 80\% of Rubin's total time, surveying almost half the sky (\SI{19600}{deg^2}). The first year will cover the sky uniformly. Since the same patch of the sky won't be revisited very often, these supernova light curves will be too sparsely sampled for reliable light-curve fitting. This is one of the main reasons a rolling cadence was suggested \citep{LSST2017, Lochner2018, Lochner2022}. \citet{Leistedt2025} then proposed and implemented a uniform rolling cadence, combining the benefits of rolling with uniformity being reached at key milestones.

The rolling cadence will start in Rubin's second year (2027), coinciding with the extension of DESI. LSST observations are taken in pairs of filters $u,g,r,i,z,Y$, with \SI{30}{s} exposures, and the ``high-activity'' regions receive revisits in a different pair every three days on average\footnote{See \url{https://survey-strategy.lsst.io/baseline/wfd.html} for observing statistics.}. High and low activity regions alternate sky locations per season (yearly), covering roughly half of the WFD each. Here, we only simulate the high-activity WFD as it is more worthwhile to spectroscopically follow up SNe that are expected to have well-sampled light curves.

\subsection{Filters, cadences, exposure times} \label{sec:filtchoice}

\begin{figure}[]
    \centering
    \includegraphics[width=0.49\textwidth]{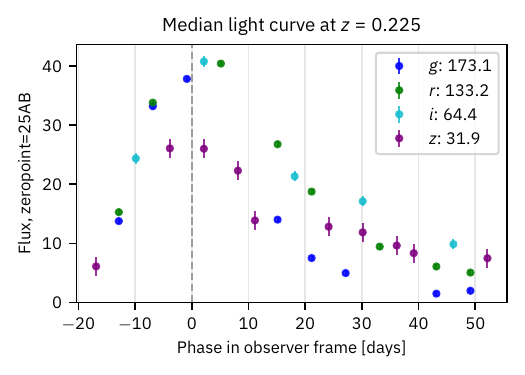}
    \caption{Simulated light curve for a median Type Ia supernova at $z=0.225$, reflecting the (high-activity) WFD strategy to observe in pairs of filters every three days. These results are similar to other simulations, e.g. fig. 16 of \citep{Kessler2019PLAsTiCC}. Missing points reflect bright moon phases (affecting $g, r, i$ predominantly), the possibility of a supernova being missed due to chip gaps, and $u$ and $Y$ bands being excluded in our cosmology analysis. The numbers in the legend are the quadrature-summed signal-to-noise values per filter.
    \label{fig:median_LC}}
\end{figure}
We emulate the high-activity WFD by assuming observations in the following filters, three days apart, and on a 28-day (lunar) cycle: $g+r, i, g+r, z, g+i, z, r, z, z$. Every observation has \SI{30}{s} exposures. For the time being, we do not include $u$-band because light curve models like SALT \citep{Guy2007, Kenworthy2021}) do not model this band well \citep{Nordin2018}, or $Y$-band because the high sky brightness results in low signal-to-noise ratios. 

In total, this amounts to two observations in $i$, three observations in $g, r$, and four observations in $z$ per lunar cycle. These cadences incorporate the effect of the Moon: Bright moon phases will have an impact on observations. Due to Rayleigh scattering, bluer light is predominantly scattered in the sky, heavily increasing the noise especially in $g,r,i$. Thus, there is a window each lunar cycle in which only $z$-band information enters our analysis.

We randomly dropped one additional observation in $i$ per lunar cycle to simulate possible light curve gaps if the supernova happens to fall into a chip gap. The omission of 1 in 13 observations is based on the fill factor of 93\% \citep{LSSTScienceBook}, and the $i$ band was chosen arbitrarily.

We found that changing the specific sets of observations has minimal impact on cosmological constraints, especially when only taking into account the relatively low-redshift supernovae that we aim to spectroscopically follow up. Because the high-activity WFD achieves high total signal-to-noise ratios for these nearby targets anyway, more densely sampled light curves improve constraints only marginally; cycling through $g, r, i, z$ every four days (while still omitting $g, r, i$ during bright moon) improves FoM values by less than 1\%. Consequently, occasional data losses due to weather also have little impact.

We assume depths where a photometric $\rm SNR>5$ (``$5\sigma$ depths'') is achieved for $g=24.50$, $r=24.03$, $i=23.41$, and $z=22.74$ for a nominal \SI{30}{s} WFD exposure. These numbers\footnote{\url{https://pstn-054.lsst.io/}} take into account real observing conditions, such as a simulated median $r$-band seeing of \SI{1.05}{\arcsec}, and are thus much worse than the higher numbers of $25.0,24.7,24.0,23.3$ listed on the overview website,\footnote{\url{https://rubinobservatory.org/for-scientists/rubin-101/key-numbers}} which correspond to idealized observing conditions (e.g. $r$-band seeing of \SI{0.70}{\arcsec}).

\subsection{Light curve and cosmology simulation}

We use the simulation code\footnote{\url{https://github.com/rubind/wfirst-sim}, used only with modest changes, adding features for ground-based surveys.} from \citet{Rubin2025} to simulate a set of supernova light curves for given survey specifications and perform the cosmology fit.

\citet{Rubin2025} implement a Fisher matrix code which emulates a UNITY-like treatment of systematics \citep{Rubin2025Union3}. We fit for distance moduli and nuisance parameters simultaneously -- at the level of individual light curve points. This takes into account all correlations between these parameters, and fully propagates their uncertainties to the distance modulus covariance matrix. 

From there, we obtain a Fisher matrix (\Cref{eq:FisherSN}) and can thus evaluate the performance of the simulated surveys. Here, we give a brief overview of the assumed uncertainties. We refer the reader to \citet{Rubin2025} and Appendix \ref{sec:Simulations} for more details.

Supernovae are generated for a given survey footprint based on volumetric rates from \citet{Rodney2014}. Individual light curves are generated with SALT3 \citep{Kenworthy2021} with a population fit to \citep{Betoule2014JLA}. Noise is added based on the previously cited $5\sigma$ depths.

\subsubsection{SN-to-SN variability}

Through light curve fits \citep{Guy2007,Kenworthy2021}, SNe~Ia can be standardized with a root-mean-square accuracy of \SI{0.145}{mag} \citep{Rigault2025ZTFDR2}. We split this into a ``gray'' dispersion of \SI{0.1}{mag} and a ``color'' dispersion of \SI{0.03}{mag} \citep{Pierel2022}. With a color standardization parameter of $\beta\approx3.5$ \citep{Rubin2025Union3}, these two dispersions combine back to the familiar $\sqrt{0.1^2+(0.03\cdot3.5)^2}\,\si{mag}=0.145\,\si{mag}$. Gray dispersions is applied as a correlated/systematic. The color dispersion is correlated per filter per supernova.

The above numbers may be conservative: Lower dispersion values are achievable with improved photometric calibration of optical (\citet{Rubin2025Union3}, Fig.~6) or with the addition of near-infrared data \citep{Pierel2022}\footnote{We included a two-week Roman-like survey as placeholder data to calibrate the SALT3-NIR model from \citet{Pierel2022}. No near-infrared data is assumed for Rubin SNe.}. We explore the impact of the chosen gray dispersion on cosmology in \Cref{fig:var_systematics}, while also showing that our main results are robust to larger values.

Host galaxy dust extinction lowers the signal-to-noise ratio, correlated per supernova (similar to the gray dispersion). 
For this, we use the \citet{Cardelli1989} color law with a standard $R_V=3.1$.

\subsubsection{Calibration and supernova model}

We assume the following Gaussian prior uncertainties on instrument calibration uncertainties:

\begin{itemize}
    \item \SI{5}{mmag} zero-point calibration uncertainty per filter, correlated across all SNe.
    \item \SI{5}{\angstrom} wavelength uncertainty per filter, correlated across all SNe.
    \item \SI{7}{mmag/\micro\metre} tilt in flux calibration vs. wavelength \citep{Bohlin2014}, correlated across all SNe
\end{itemize}

Uncertainties in the trained supernova model are also taken into account. This is an important step, as ignoring model uncertainties leads to overly optimistic cosmology forecasts (see \citealt{Rubin2025} and Appendix \ref{sec:Simulations}).

The Bayesian fit typically finds posterior calibration uncertainties smaller than their priors. This is known as self-calibration \citep{Kim2006}: The behavior of an average SNe~Ia is assumed consistent across different redshifts. That is, a higher-redshift supernova should behave similarly in a redder filter when compared to a lower-redshift supernova in a bluer filter. If the calibration of one filter is off, the model that best fits all supernovae simultaneously will detect that inconsistency — and adjust the calibration parameters accordingly.

In contrast, frequentist analyses \citep{Brout2022Pantheon, Popovic2025b} self-calibrate to a much smaller extent due to the lack of the simultaneous analysis of all light curves: Calibration parameters are uncorrelated by construction, and the final covariance relies entirely on the assumed ``prior'' uncertainties.

\begin{figure}[]
    \centering
    \includegraphics[width=0.49\textwidth]{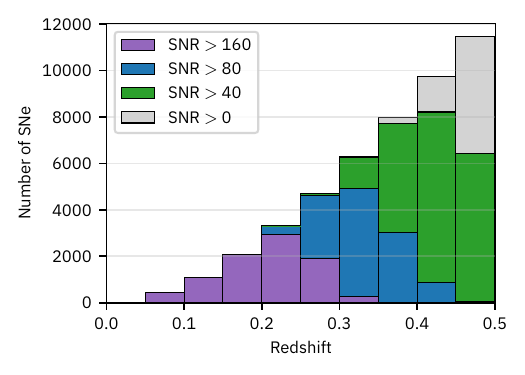}
    \caption{Histogram of forecasted SNe~Ia discovered in our simulations of Rubin's high-activity Wide Fast Deep in the overlap region with DESI in one year. Practically all SNe are well-measured up to $z=0.35$. The SNR values given here are defined to be the quadrature sum of all measurements of a single supernova. 
    \label{fig:redshift_hist}}
\end{figure}

\subsection{Redshift completeness}

Redshift completeness refers to the ability of a survey to capture the true population of supernovae up to a certain redshift. Given the magnitude limits of Rubin, we want to estimate below which redshift the sample is redshift-complete, i.e., few intrinsically fainter SNe are lost (Malmquist bias, \citet{Malmquist1925}).

In the redshift bin $z=0.3...0.35$, over $99.7\%$ of SNe have a combined signal-to-noise ratio of over 40 (\Cref{fig:redshift_hist}), and are thus considered well-measured by the simulation. It is thus highly improbable that the Rubin WFD loses intrinsically fainter normal SNe below $z_{\rm WFD}=0.35$.

Because Rubin provides an essentially complete and unbiased parent catalog up to this redshift, the limiting factor for our final volume-limited cosmology sample is dictated entirely by our follow-up capabilities. We thus investigate this spectroscopic completeness, i.e., the ability of DESI to observe this complete sample, separately in \S\ref{sec:DESI}.


\section{DESI spectroscopic follow-up} \label{sec:DESI}

\begin{figure*}[]
    \centering
    \includegraphics[width=0.99\textwidth]{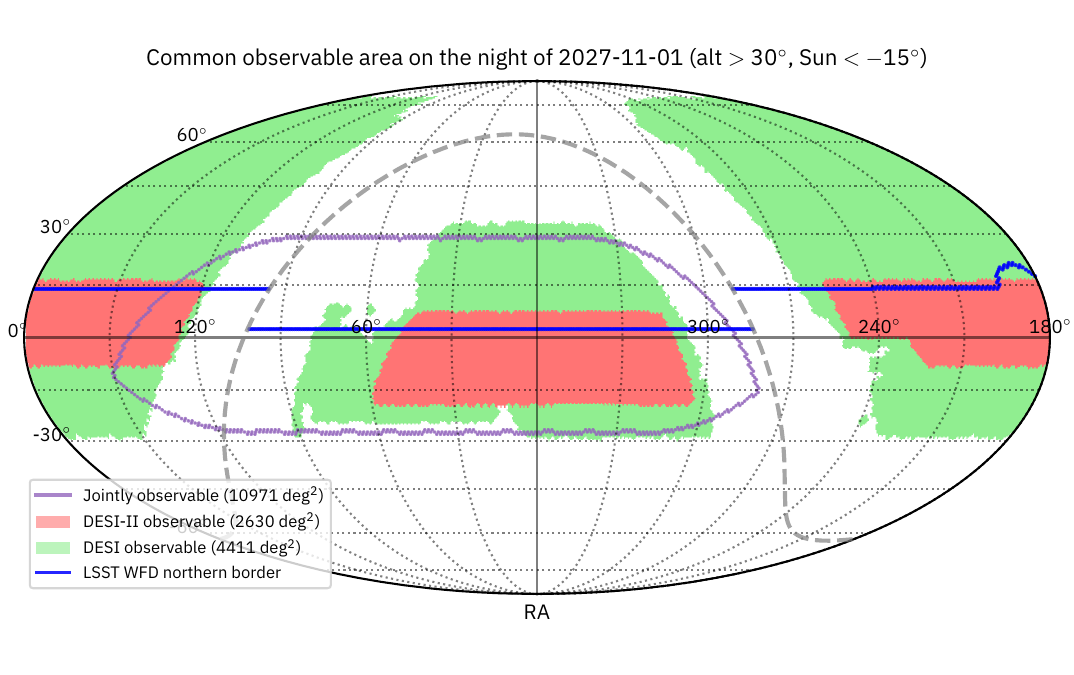}
    \caption{DESI, DESI-II and approximated LSST Wide Fast Deep footprints. The numbers in the legend refer to the areas inside DESI(-II) and LSST-WFD that are observable by both sites in a single night. The total footprint overlaps with LSST-WFD are \SI{9000}{deg^2} for DESI and \SI{5000}{deg^2} for DESI-II.
    \label{fig:footprint}}
\end{figure*}

\begin{figure*}[]
    \centering
    \includegraphics[width=0.99\textwidth]{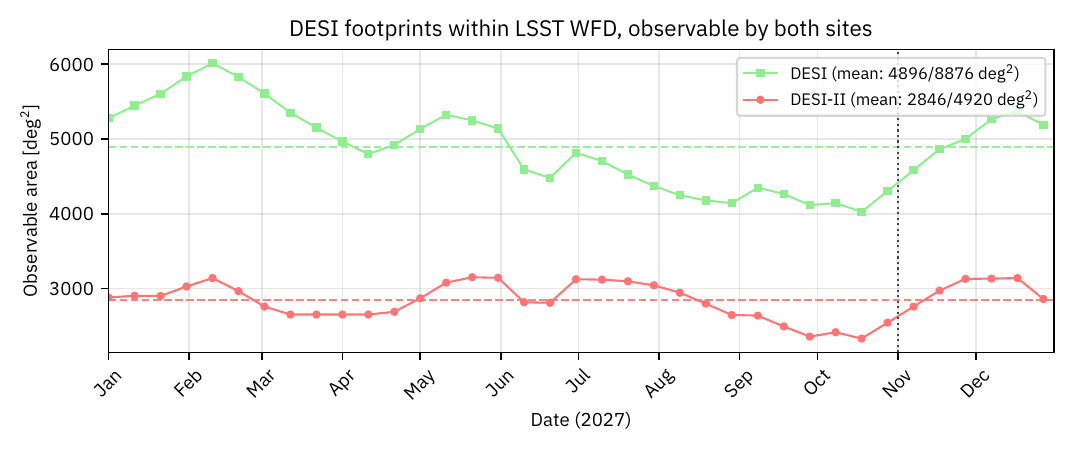}
    \caption{The nightly observable area of DESI and DESI-II footprints within LSST-WFD in 2027, requiring object altitude $>30^\circ$ and Sun altitude $<-15^\circ$ at both sites.
    \label{fig:observable_area_2027}}
\end{figure*}

The Dark Energy Spectroscopic Instrument uses its multi-fiber spectrograph to observe up to 5000 spectra per pointing, in a range from 3600 to \SI{9800}{\angstrom} \citep{DESI2016, Poppett2024}. DESI has had its initial survey extended to cover a footprint of at least \SI{17000}{deg^2} by the end of 2028.

DESI uses the 4-meter Mayall telescope at Kitt Peak National Observatory (northern hemisphere), while Rubin is on Cerro Pachon in Chile (southern hemisphere). Naturally, there will be limited overlap in survey footprints. But having covered the northern sky, the extension of  DESI specifically focuses on tiles further to the south, increasing the overlap region with Rubin's WFD (see \Cref{fig:footprint}).

Ideally, one would like to spectroscopically follow up all supernovae discovered by Rubin. But devoting a chunk of DESI's time fully to supernovae would be a suboptimal use of the instrument. No more than a few supernovae would be measured per pointing (DESI's FoV is \SI{8}{deg^2}), while the instrument has 5000 fibers. Thus, an efficient DESI supernova program should be embedded into the main survey, only borrowing a couple of the free fibers whenever the opportunity arises.

DESI's main survey consists of a ``dark'' and ``bright'' program (see \citet{Schlafly2023} for survey operations). Dark tiles are observed during dark time, and the ``bright'' program is observed when conditions are worse due to the impact of sky brightness, seeing and atmospheric transparency. Dark and bright tiles receive effective exposures time of ($t_{\rm eff}=\SI{1000}{s}$) and ($t_{\rm eff}=\SI{180}{s}$) respectively (or \SI{16.7}{min} and \SI{3}{min}), so the dark tiles are intended to observe much deeper than the bright tiles. Effective exposure time refers to an observation under nominal dark conditions (\SI{1.1}{\arcsec} seeing in r-band, r-band sky brightness of 21.07 AB mag, no clouds, and pointing at zenith). When observing, DESI takes the actual conditions into account in real-time, and calculates the necessary ``wall-clock'' time such that each pointing receives a similar depth. Because the Moon can impact SNRs heavily, the wall-clock time per exposure is often similar for dark and bright tiles.

We assume that the main program is allotted 670~hours of dark time and 110~hours of bright time per year (\citet{Schlafly2023} and D. Schlegel, priv. comm.). Each location on the sky receives an average of 5 visits over multiple tiles, leading to a uniform coverage of the whole survey area.

The current DESI Transients Survey program \citep{Hall2026} operates with what we call a ``passive trigger'': DESI proceeds with pre-planned pointings, and SNe only trigger the usage of spare fibers within a planned pointing, i.e., they are only observed if they happen to fall into a pre-planned pointing, leaving no control over when the follow-up is scheduled.

Instead, we suggest that as long as a DESI tile still has remaining visits left, Rubin-discovered transients may trigger one of these visits. We call this an ``active trigger'', and aim to take spectra near peak brightness. Active triggers dynamically reorder the queue, allowing many more live supernovae to be followed up near peak brightness -- and at little-to-no additional cost to DESI's main program.

In this section, we evaluate the performance of such a one-year integrated DESI supernova spectroscopy program during Rubin's year 2, when the WFD rolling cadence commences: How good is DESI at spectroscopically following up live SNe, obtaining redshift measurements, and classifying them?

\subsection{Depths at nominal effective exposure times}

To begin, we estimate the maximum redshifts at which one can reliably distinguish SNe~Ia from other transients with one nominal dark ($t_{\rm eff}=\SI{1000}{s}$) or bright ($t_{\rm eff}=\SI{180}{s}$) exposure. We call the redshifts to which we can obtain a volume-limited spectroscopic sample $z_{\rm VL}^{\rm dark}$ and $z_{\rm VL}^{\rm bright}$ respectively.

In general, it is preferable to observe a supernova near peak, because the signal is stronger and spectral features are more distinct \citep{Branch2006}. When simulating the observation scheduling, we require the candidate to be $\pm5$~days within peak. This is chosen based on our tests showing SNR drops off no more than 10\% when observing at $\pm5$~days from maximum (in the observer frame) relative to peak brightness for redshifts around 0.3.

One must avoid spectroscopic selection effects, since it is much easier to spectroscopically classify intrinsically brighter supernovae. When determining the highest ``classifiable redshift'', we thus require that supernovae intrinsically fainter by $\Delta M=\SI{1.0}{mag}$ be classifiable. This covers $>99.9\%$ of SNe in populations similar to \citet{Betoule2014JLA}, meaning that we would indeed obtain a redshift-complete spectroscopic sample representative of the underlying population.\footnote{This drops to $99.5\%$ when using the standardization parameters from \citet{Rubin2026Union3.1}.}

To determine whether a SN~Ia is classifiable, we use visual inspection of the characteristic SiII$\lambda 6335$ feature as an initial check. This feature is Dopper shifted by the photospheric expansion to \SI{6150}{\angstrom} in the supernova's rest frame \citep{Filippenko1997}. Rubin's photometric redshifts allow us to know which wavelengths we should examine for this and other spectral features. This SiII feature is within the wavelength range of DESI up to $z\sim0.6$.

We also compute the signal-to-noise ratio per \SI{15}{\angstrom} from 4500-8000\,\si{\angstrom} (hereafter denoted $\rm SNR_{15\si{\angstrom}}$) to compare to values in \citet{Frohmaier2025TiDES} (TiDES). They found that a minimum $\rm SNR_{15\si{\angstrom}}$ in the range of 3 to 5 is sufficient to reliably classify supernovae, including sub-types, even without a redshift prior.

\subsubsection{Generating noisy spectra}

\begin{figure*}[t]
    \centering
    \includegraphics[width=0.99\textwidth]{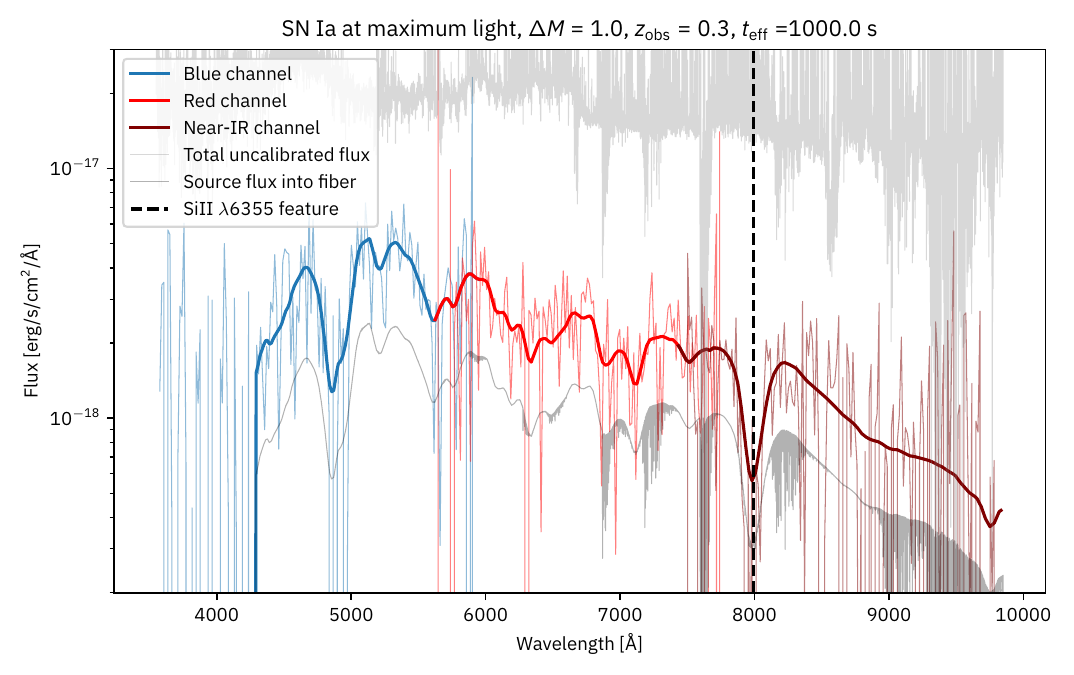}
    \caption{A faint Type Ia supernova spectrum at maximum light (bold colored lines), with added noise (thin colored lines) as simulated by DESI's \texttt{SpecSim} for a single dark exposure. Binned in \SI{15}{\angstrom}. $\rm SNR_{15\si{\angstrom}}=3.0$, total $\rm SNR = 52$.
    \label{fig:desi_snia}}
\end{figure*}
\begin{figure}[]
    \centering
    \includegraphics[width=0.49\textwidth]{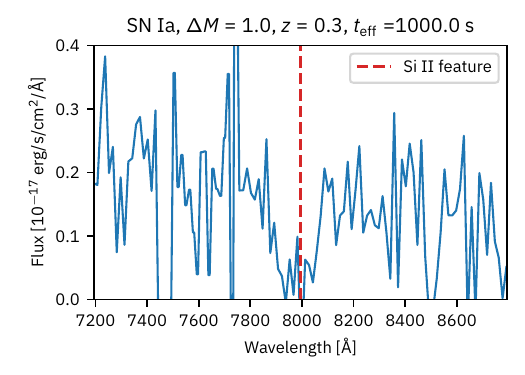}
    \caption{The same simulated SN~Ia spectrum as in Figure~\ref{fig:desi_snia}. Even without the visual cue from the template, the SiII absorption line is still readily visible.
    \label{fig:desi_snia_zoom}}
\end{figure}

To simulate realistic spectra, we use DESI's \texttt{SpecSim}\footnote{\url{https://github.com/desihub/specsim}} and feed into it a SN~Ia with average intrinsic properties as generated by the Probabilistic Autoencoder (PAE) from \citet{Stein2022}, which was trained on spectral time series from the Nearby Supernova Factory \citep{Aldering2002SNFactory} with a rest frame wavelength range from 3300 to 8600\,\si{\angstrom}. The PAE is constructed such that setting all input parameters to 0 generates spectral time series of an average supernova. In order to test behavior away from maximum light, we generate a total of 32 spectra, one day each from $-10$ to $+21$~days in the supernova's rest frame.

Then, \texttt{SpecSim} generates noise at the level of number of electrons\footnote{\url{https://specsim.readthedocs.io/en/stable/api/specsim.simulator.Simulator.html\#specsim.simulator.Simulator.generate_random_noise}}, including Poisson noise from the source (SN), sky, dark and Gaussian read noise, before propagating that to noise flux in the measured signal. We use standard settings, including an airmass of 1 and no additional noising due to suboptimal observing conditions, so that the input exposure times are indeed equivalent to the effective exposure times.

To simulate a faint supernova, we for now, and for the sole purpose of estimating $SNR_{15\AA}$ near the SiII~6150\AA feature, simply employ a gray (wavelength-independent) offset of the aforementioned $\Delta M=\SI{1.0}{mag}$, which dampens the flux by 60\%. The gray offset leaves open the question whether such a faint supernova is intrinsically faint, or appears faint due to dust extinction. If the latter is the case, for the same $\Delta M$, bluer wavelengths would be affected more strongly than wavelengths around the critical SiII line. So such a gray offset is a conservative estimate.\footnote{We explicitly compare synthetic photometry to an \texttt{SNCosmo}-generated unstandardized light curve of a SN with parameters $(x_1,c)=(-2,0.2)$, which is $\sim\SI{1}{mag}$ fainter in $m_B$, and find that the gray offset imposed here is conservative by $\sim\SI{0.1}{mag}$ in $r$-band, and $\sim\SI{0.2}{mag}$ in $i$-band, which is where the SiII line will be redshifted.}

\subsubsection{Results and discussion}

We find that with a single DESI dark time effective exposure of \SI{1000}{s}, one can reliably classify (99.9\% complete) SN~Ia populations up to $z_{\rm VL}^{\rm dark}=0.3$ via the SiII$\lambda 6335$ feature alone. An example spectrum of a faint SN~Ia ($\Delta M=\SI{1.0}{mag}$) at $z=0.3$ is given in \Cref{fig:desi_snia}, with a zoom-in on the SiII feature in \Cref{fig:desi_snia_zoom}. That spectrum has a corresponding $\rm SNR_{15\si{\angstrom}}=3.0$. In comparison, an average-brightness supernova ($\Delta M=\SI{0.0}{mag}$) has $\rm SNR_{15\si{\angstrom}}=7.2$.

Observing the same faint $z=0.3$ supernova with one dark exposure at $+5$ or $-5$~days (observer frame) from maximum light leads to a drop to $\rm SNR_{15\si{\angstrom}}=2.7$, with the SiII feature still visible.\footnote{This is important to verify, even though practically, an edge case like this will rarely come into play: One will always schedule around the faintest supernova, trying to capture it at maximum light. So the only scenario where this happens is when multiple faint SNe close to $z=0.3$ with different peak times happen to be in the same tile at the edges of one 10-day observation window. And in this rare case, it might be preferable to simply schedule two separate pointings. Coincidentally, SNe at higher redshifts are also more forgiving in this regard -- they stay near peak for longer in the observer frame by the $1+z$ time dilation factor.} 

Going to $z=0.35$ leads to a significant reduction in the SNR as anticipated, with $\rm SNR_{15\si{\angstrom}}=2.1$ for a faint SN~Ia ($\Delta M=\SI{1.0}{mag}$) for $t_{\rm eff}=\SI{1000}{s}$ in dark time. Although the SiII$\lambda 6335$ feature is noticeably more noisy, classification could still be possible when additionally taking into account other spectral features. Classifiability even without a visually discernible SiII line was shown to be possible in \citet{Hall2026}. This also indicates that the \citet{Frohmaier2025TiDES} criterion of $\rm SNR_{15\si{\angstrom}}>3$ tends to be conservative -- if a photo-$z$ prior exists. Nevertheless, we err on the safe side and do not include $z>0.3$ SNe in our analysis, rather using the fact that lower-SNR SNe are likely classifiable \citep{Hall2026} as margin. Future work is needed to investigate a more exact classifiability criteria for our case, where a volume-limited sample is desired to reduce selection effects.

For a single bright exposure of \SI{180}{s}, we determine a maximum redshift of $z_{\rm VL}^{\rm bright}=0.15$. Here, a faint SN~Ia has $\rm SNR_{15\si{\angstrom}}=3.7$. An average SN~Ia has $\rm SNR_{15\si{\angstrom}}=8.6$. Going from $z=0.15$ to $z=0.2$ drops $\rm SNR_{15\si{\angstrom}}$ to $1.9$ for a faint SN~Ia, with classifiability being unlikely.

We emphasize that we use the SiII$\lambda 6335$ feature more as a check for classifiability; one should not solely rely on it. For example, at redshift $z=0.24$, the SiII feature and the atmospheric $\rm O_2$ Fraunhofer A absorption line at \SI{7600}{\angstrom} (visible in Figure~\ref{fig:desi_snia}, especially for the ``source flux into fiber'') overlap, leading to a much more noisy SiII$\lambda 6335$ feature. In this case, however, SNRs are high with one dark time exposure, with $\rm SNR_{15\si{\angstrom}}=10.0$ for an average SN~Ia and $\rm SNR_{15\si{\angstrom}}=4.2$ for a faint Ia, so we find that the feature is still visible. But even if not, one can rely on the other (well-measured) features to classify.

To summarize, DESI can provide reliable spectroscopic classification for a volume-limited sample out to $z_{\rm VL}^{\rm dark}=0.3$. Although this does not reach the optimal redshifts at $z=0.5$ to $0.6$ found in \S\ref{sec:toymodel}, \Cref{fig:z_1d_grid+lowz} shows that there is still a significant amount of cosmological information at $z<z_{\rm VL}^{\rm dark}=0.3$.

\subsection{Scheduling spectroscopic follow-ups}

\begin{figure}[]
    \centering
    \includegraphics[width=0.49\textwidth]{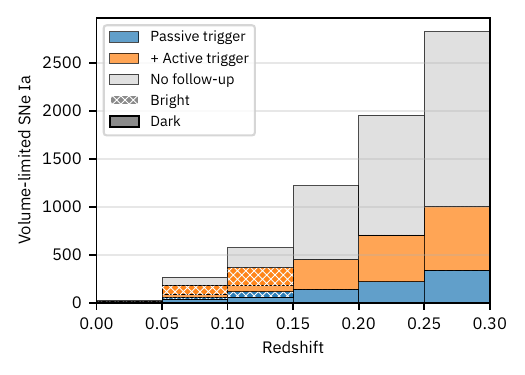}
    \caption{Stacked histogram showing the one-year forecast of spectroscopically followed near-peak ($\pm 5$~days) volume-limited SNe~Ia per redshift. The current DESI Transients Survey program, where spare fibers are assigned to live transients if they happen to fall into a planned pointing (blue), would obtain around 880 SNe~Ia. When allowing a transient to actively trigger any remaining tile pointing, 2300 SNe~Ia can be followed up near peak (orange). This does not impact the completion of DESI's main survey as SN candidates only the reorder the observing queue. This is effectively a two-tier survey: We have a wide tier that can cover a larger total area with shorter exposure times (bright tiles, $z<0.15$), and a deep tier that covers a smaller total area with longer exposure times (dark tiles, $z<0.3$).
    \label{fig:sne_per_redshift}}
\end{figure}

To study the yield of a practical follow-up plan, we generate transients in the high-activity region of Rubin's Wide-Fast-Deep and implement a sliding window in which DESI could follow up these transients (\Cref{fig:footprint}). Candidates are generated with a ten-day lifetime in which they can be followed up, and after which they expire -- this aims to get them observed near peak consistently. Every night, tiles are ranked by the number of ``alive'' objects. The top-ranking tiles are then allowed to be ``actively triggered'' as long as said tile has remaining passes left. \Cref{fig:sne_per_redshift} summarizes the number of SNe~Ia we are able to follow up with this scheme. In the following, we explain this process.

\subsubsection{Volume-limited SNe~Ia and ``noise''}
\label{sec:snia_vs_non}

DESI will rank tiles based on the total number of objects that pass a magnitude cutoff -- without knowledge whether it is a desired object, i.e., a SN~Ia in the volume-limited sample below $z_{\rm max}$, or noise, i.e., non-Ia transients and other SNe~Ia. This can lead to tiles with fewer desired SNe~Ia being scheduled over tiles with more desired SNe~Ia, an effect that impacts DESI's proficiency at obtaining a volume-limited sample of SNe~Ia.

Here, we estimate the probability $p$ that an object in a magnitude-limited survey is indeed a SN~Ia belonging to the volume-limited sample. 

The Zwicky Transient Facility found that over 70\% of transients were SNe~Ia after basic quality cuts, corresponding to $N_{\rm tot}/N_{\rm Ia}\approx1.39$ \citep{Perley2020}. Similarly, DES-5year finds $N_{\rm tot}/N_{\rm Ia}\approx1.35$ after some basic filtering \citep{Moeller2022}. With a simple toy model calculation for a magnitude-limited survey (Appendix \ref{sec:SNrates}), we find $N_{\rm tot}/N_{\rm Ia}\approx1.28$. Here, we will assume non-Ia objects at a ratio $N_{\rm tot}/N_{\rm Ia}=4/3\approx 1.33$.

Early-phase-only classifiers exist \citep{Qu2022, DeSoto2024}, but accuracies are only around 85\%, so they are not a suitable candidate selection tool for our purposes here: While false positives (non-Ia appearing as Ia) will be disregarded after spectroscopy, false negatives (Ia gets wrongly excluded) may have a systematic cause, biasing the final cosmological sample of SNe~Ia.

Similarly, we do not rely on photometric host galaxy redshifts to exclude SNe~Ia above $z_{\rm max}$ from contributing to a tile's score in this analysis. To estimate the fraction of SNe~Ia below and above $z_{\rm max}$, we compute the simulated signal-to-noise cut with which the population would be 99.9\% complete below $z_{\rm max}$. This SNR cut is translatable to a magnitude cut that has yet to be determined.

Then, we calculate how many SNe that in total satisfy the SNR cut are below that redshift cut. The fraction of SNe~Ia below the redshift cuts are $0.42$ for both the $z_{\rm VL}^{\rm bright}=0.15$ and $z_{\rm VL}^{\rm dark}=0.3$ cuts. In comparison, ZTF found a value of $0.37$ \citep{Rigault2025ZTFDR2}. In this study, we assume a fraction of $0.4$. 

Combined with the non-Ia fraction, the probability of an object being a ``desired'' volume-limited SN~Ia is $p=0.3$.

Small percent-level changes in $p$ lead to even smaller changes in the resulting volume-limited sample, and the general effect of $p$ on the follow-up efficiency is discussed in future work (J. Truong et al., in prep).

\subsubsection{DESI tile availability}

DESI observes the same location on the sky multiple times with different, slightly-offset tiles. We want to take advantage of that, actively triggering a tile visit if it contains many potential SNe~Ia. 

For simulation purposes, we tile the total DESI/WFD high-activity footprint overlap into discrete, non-overlapping tiles of \SI{8}{\deg^2} (DESI FoV). How many pointings remain for each simulated tile is calculated based on the actual remaining DESI tiles. The numbers are different for the bright and dark tiles.

Each DESI pointing has a fill factor of 70\% of the \SI{8.0}{deg^2} field-of-view that is accessible from the active fibers. Since the number of supernovae per pointing is of the order of one, this should not lead to losses as long as a dithering scheme is implemented that guarantees good fibers for all candidates within a tile.

\subsubsection{Observability of generated supernovae}

Transients are generated directly into the discretized tiles described earlier, which span the entire Rubin high-activity WFD and DESI overlap region. Each tile independently generates volume-limited SNe~Ia based on redshift-dependent rates \citep{Rodney2014}, along with ``noise'' at the fraction discussed above. We determine the (time-dependent) area that DESI can follow up as follows (\Cref{fig:footprint}).

The (instantaneous) observable area is computed given the locations of the telescopes, while requiring the observed object to be (i) in both WFD and DESI footprints, (ii) at an altitude $>$\SI{30}{\deg} above, and (iii) the Sun to be at $<$\SI{-15}{\deg} below the horizon. The daily observable area is then the intersection of all instantaneous observable areas over one night (\Cref{fig:footprint}). We compute the mean from values calculated 10~days apart and spread over the whole year. The mean daily observable area is $4900\pm540$ \si{\deg^2} for DESI (and $2850\pm240$ \si{\deg^2} for DESI-II, a potential upgrade of DESI that would start in 2029).

Year two of Rubin is when its rolling cadence commences, and it observes the designated high-activity region for the whole year, then high- and low-activity regions swap for the following season\citep{Lochner2018}. We estimate the percentage of the general DESI/WFD overlap to be 60\% high-activity in 2027.\footnote{The 60\% number is based on \url{https://survey-strategy.lsst.io/baseline/wfd.html}, excluding deep drilling fields. The number of SNe~Ia scales to first order with $\sqrt{2\ln M}$, where $M$ is the number of tiles we can choose from (J. Truong et al., in prep), so forecasts only depend weakly on this percentage.}

\subsubsection{Active trigger: Scheduling follow-ups}

The observable area constraint is implemented as a sliding window within which DESI tiles may be triggered if they have not been observed yet. We allow a burn-in phase of ten days, which is also our defined lifetime of an object. After that, tiles with the most candidates in them are scheduled each night if they have DESI visits remaining. The effective time allotted to the main survey corresponds to six bright and seven dark exposures per night on average.

This procedure emulates a real observing pipeline: Before each night, one evaluates a stream of potential candidates coming from Rubin. Dark tiles receive a score based on how many candidates pass the dark magnitude cut, and the top-ranking dark tiles are scheduled. The scheduled candidates are removed from the queue, after which the top-ranking bright tiles are scheduled -- based on a different magnitude cut for the bright tier. Dark tiles are scheduled first to emulate a true wide/deep wedding-cake style structure and to reduce selection effects (J. Truong et al., in prep).

This simulated follow-up program runs for a total of 320~days. This accounts for the fact that we won't follow up SNe with incomplete light curves -- these SNe would have peaks near the beginning and end of the survey.\footnote{We obtained this 320 day window from the output of the simulation described in \S\ref{sec:sims}; it is the difference between the earliest and latest day of peak of a supernova taken into account in the cosmological analysis.} We use redshift bins with 0.05 width, starting from $z=0$, even though new DESI-Rubin SNe below $z=0.05$ are not included in the cosmological analysis (see Appendix \ref{sec:Simulations}).

\subsubsection{Follow-up results}

\begin{figure*}[]
    \centering
    \includegraphics[width=0.99\textwidth]{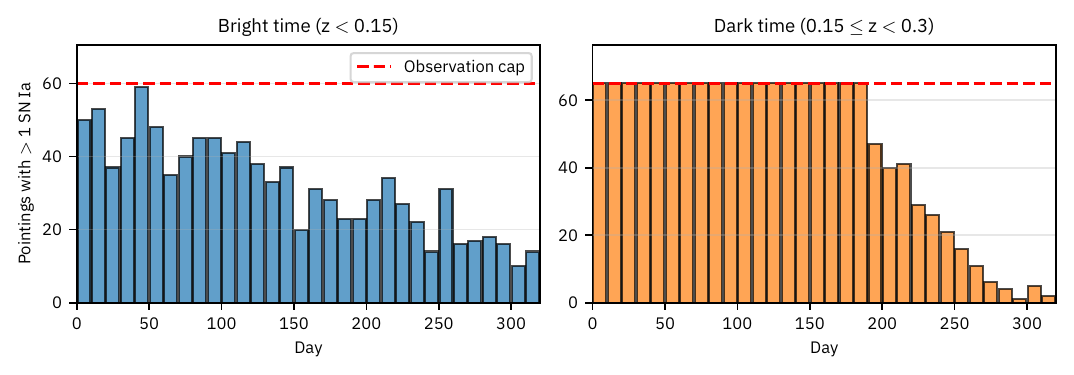}
    \caption{Number of pointings with at least one near-peak ($\pm 5$~days) SN~Ia per 10-day bin for bright (left) and dark (right) tiles, compared to the available observing time (red). For bright tiles, the small volume at $z < 0.15$ means that available tile pointings initially exceed the supernova rate, but efficiency still decreases continuously as tiles are depleted. For dark tiles, the larger volume produces more SNe, so the program is initially limited by the possible observation time rather than by available tiles or supernova counts. After roughly half a year, the depletion of tiles in the DESI/high-activity-WFD overlap region becomes the dominant bottleneck, limiting the number of SNe that can be followed up.}
    \label{fig:bright_dark_tiles_used}
\end{figure*}

 This baseline survey is defined by the currently 
remaining DESI tiles, meaning that these spectroscopic observations would come at no added effective exposure times -- all we do is allow SNe to trigger pointings that will eventually take place anyway.

With active triggers, 7500 Rubin-discovered objects will be spectroscopically followed up near peak brightness ($\pm 5$~days) in total. 5600 will be confirmed as SNe~Ia, out of which 2300 belong to the volume-limited sample. The bright program would discover 390 volume-limited ($z<z_{\rm VL}^{\rm bright}=0.15$) SNe~Ia, and the dark program gets 1900 volume-limited ($z<z_{\rm VL}^{\rm dark}=0.3$) SNe~Ia.

In contrast, DESI would only follow up 880 volume-limited SNe~Ia with a ``passive trigger'' approach, where spare fibers are assigned if the opportunity arises \citep{Hall2026}, but tile visits are generally scheduled without considering live transients. This number assumes that tiles can be dithered to guarantee good fibers for all candidates, but this is not what is currently implemented. The current approach, if unchanged, would actually reduce the ``passive trigger'' yield to $\sim 600$ SNe~Ia. 

We find that active triggers produce much better yields, especially for bright tiles. The number of volume-limited low-$z$ SNe~Ia is increased by a factor of three, making this approach highly effective at collecting a large spectroscopically confirmed low-redshift supernova sample.

One can also use additional spare fibers to follow up objects in an extended time window before and after maximum light if they are in an already scheduled tile -- that is, without counting them when ranking tiles. For a window of $-15$ to $+45$~days from maximum light (rest frame), a total of over 20,000 spectra could be obtained; a single object would be measured a bit more than twice on average within this expanded time window.


Many regions have tiles with only one DESI pointing left, so if multiple SNe are discovered in these regions some time apart, our baseline program would not allow the tile to be actively triggered. In our study, this significantly reduced the available sky area in which SNe can be followed up, leading to many pointings not containing any SN~Ia. For the bright tiles, this is the limiting factor right from the beginning. For the dark tiles, this becomes the bottleneck half a year into the survey (\Cref{fig:bright_dark_tiles_used}).

\begin{figure*}[]
    \centering
    \includegraphics[width=0.99\textwidth]{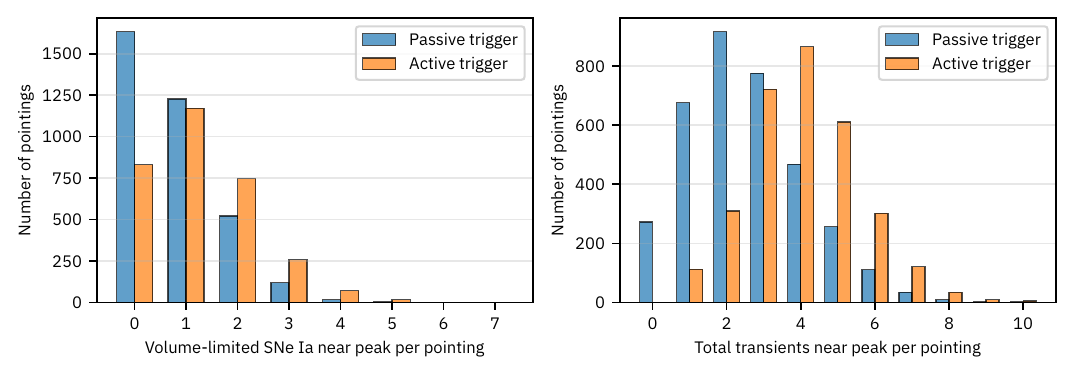}
    \caption{Efficiency: Fibers per pointing used on volume-limited SNe~Ia (left) and all candidates passing the magnitude cut (right) within $\pm 5$~days of maximum light, for bright and dark tiles combined. Even with active triggers, some pointings will not contain a volume-limited SN~Ia because tiles can only be scored and scheduled based on the number of all candidates, out of which only around 30\% belong to the volume-limited sample of SNe~Ia.}
\end{figure*}

\subsection{Spectroscopic follow-up discussion}

A DESI spectroscopic follow-up program for Rubin would generate the largest volume-limited sample of spectroscopically confirmed type Ia supernovae with well-controlled photometric systematics to date. Even when operating under the restricted tile budgets of DESI's main program, one obtains a volume-limited ($z<0.3$) sample of 2300 SNe~Ia spectra near peak in one year.

In comparison, TiDES \citep{Frohmaier2025TiDES} will spectroscopically measure 13000 SNe~Ia over its 5-year duration. However, TiDES plans to observe with fixed ``wall-clock'' exposure times instead of achieving consistent depths by multiplying an effective exposure time by a factor to compensate for non-ideal observing conditions. Also, currently TiDES does not implement active triggers. Thus, SNe~Ia are observed by chance --- similar to DESI ``passive triggers'' but with varying signal-to-noise ratios that lead to selection effects at significantly lower redshift. For TiDES it is estimated that out of the 13000 SNe~Ia, 6000 are found at $z<0.3$, amounting to 1200 $z<0.3$ SNe per year.\footnote{This number is slightly larger than the 880 per year we find with DESI passive triggers despite DESI's more consistent depths, because the TiDES footprint almost completely overlaps with Rubin, while DESI's does not.}

So TiDES not only produces fewer $z<0.3$ SNe per year than our proposed program, but their sample will also suffer from selection effects: Where a volume-limited sample has a number distribution $N(z)$ proportional to the comoving volume slice, which implies $N'(z)>0$ and $N''(z)>0$ (for each survey tier), the distribution for TiDES already begins to flatten at $z=0.15$ (\citet{Frohmaier2025TiDES}, fig. 4). This indicates that the SNe at $z>0.15$ are not representative of the underlying population; intrinsically fainter SNe are lost.



Even with the large efficiency gains by actively triggering the optimum DESI tiles, there are several avenues to improve upon the simple scheduling scheme presented here.

One could aim to enhance the fraction of useful, volume-limited SNe~Ia, which we called $p$ earlier. In principle, $p$ can be enhanced by using spectroscopic host galaxy redshifts whenever available to exclude objects above $z_{\rm max}$ in the scoring. Host galaxy photometric redshifts also could be used to down-weight SN candidates beyond the nominal volume limit. The photometric redshift error bar and confidence would need to be employed careful, to avoid lowering the score for a tile having a true $z<z_{\rm VL}^{\rm dark}=0.3$ SNe~Ia. Importantly, once such a tile is selected, a fibers can be placed on such down-weighted candidates, providing feedback on this part of the selection process.

Another possibility is to exploit the fact that repeat visits to the same sky area occur through slightly-offset tiles. Being able to choose which one of the visits should be triggered can further increase efficiency (more SNe per pointing), especially in regions where many visits per tile remain. In contrast, our current simulation assumes rigid tile locations.

Building and refining a scheduling pipeline that is both efficient (large $p$) without introducing selection effects will be an important future effort, beyond the scope of this initial exploration.

\section{Results} \label{sec:Results}

\begin{figure*}[p]
    \centering
    \includegraphics[width=0.7\textwidth]{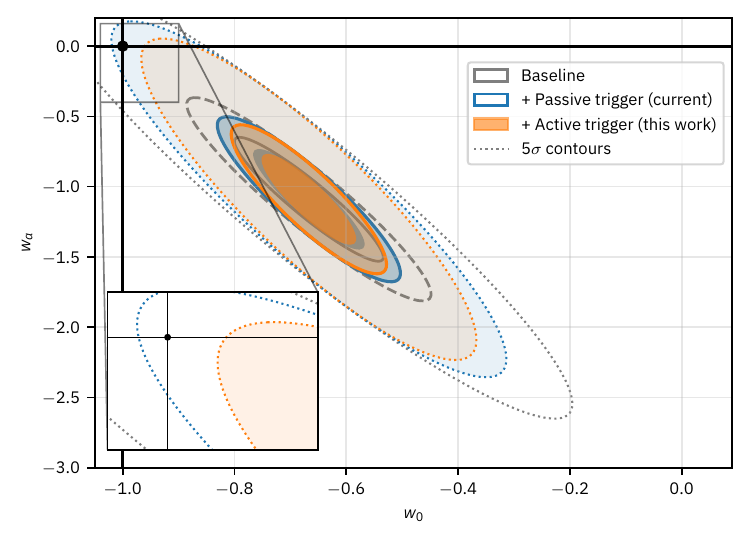}
    \caption{Projected $1, 2$ and $5\sigma$ contours for the $w_0$ and $w_a$ parameters of dynamical dark energy, and an inset zoomed in around \lcdm\ ($w_0=-1,w_a=0)$. We combine the current data baseline (gray, DESI BAO+CMB+Union3+low-z SNe as described in \S\ref{sec:toymodel}) with one year of DESI's current transient program (blue), and our proposed survey with SNe actively triggering tile visits (orange). Without increasing total effective observing time, this would lead to significantly tighter constraints, excluding \lcdm\ at $>5\sigma$ for the current best-fit point.}
    \label{fig:ellipses_DESI_SN}

    \vspace{2em}

    \begin{minipage}[t]{0.49\textwidth}
        \centering
        \vspace{0pt}
        \includegraphics[width=\linewidth]{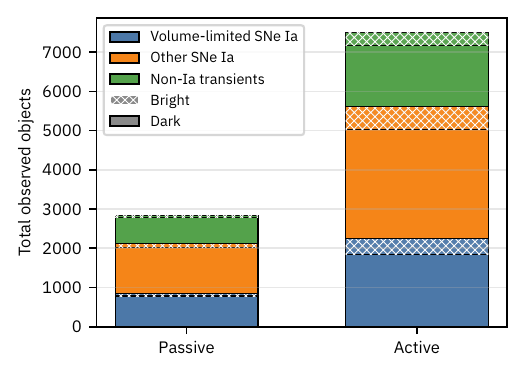}
        \captionof{figure}{Stacked histogram of DESI spectroscopic follow-up yields of Rubin WFD targets. Our proposed survey observes around 7500 total objects within $\pm5$ days of their peak brightness with 3100 actively-triggered pointings. The current spare fiber program would obtain 2800 objects with 3500 passive pointings.}
        \label{fig:total_objects}
    \end{minipage}%
    \hfill
    \begin{minipage}[t]{0.49\textwidth}
        \centering
        \vspace{0pt}
        \includegraphics[width=\linewidth]{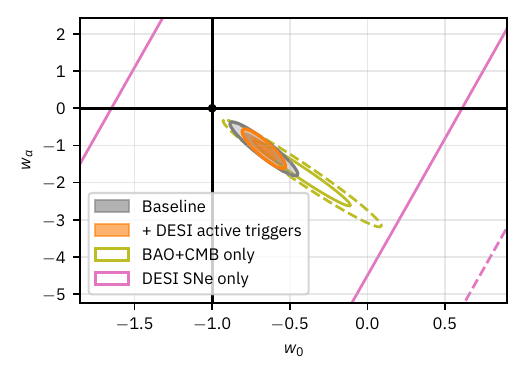}
        \captionof{figure}{$1$ and $2\sigma$ projections -- SNe only vs. BAO+CMB. The SNe-only contour is almost orthogonal to BAO+CMB. So even though the new SNe are unconstraining on their own (Figure of Merit of only 0.04), combined constraints greatly improve, with Figure of Merit gains of over 50\%. This reinforces the importance of evaluating surveys holistically, in combination with other probes.}
        \label{fig:ellipses_by_source}
    \end{minipage}
\end{figure*}

\subsection{Cosmology forecast}

This program's significant advantage arises by reordering DESI's already planned observations: Supernovae are just used to actively trigger them. Given the finite number of available DESI tile visits, not all SNe discovered in the overlapping survey area can be spectroscopically followed up. To avoid biases, we only include these spectroscopically confirmed SNe~Ia within a volume limit in our cosmology analysis. 

Our follow-up program is --- in terms of $N_{\rm SN}(z)$ --- equivalent to having a ``wedding cake''-structured survey that has a wide and a deep tier with different sky coverage for each, within which all SNe are observed and followed up. We compute that the projected 2300 total SNe are equivalent to full photometric and spectroscopic observations of a deep tier with \SI{500}{\deg^2} ($z<z_{\rm VL}^{\rm dark}=0.3$) plus a wide tier \SI{780}{\deg^2} ($z<z_{\rm VL}^{\rm bright}=0.15$) every day for a year.

The current DESI Transients Survey \citep{Hall2026} does not change the tile ordering and would only produce 880 supernovae even with optimal fiber assignment, equivalent to a wide tier of \SI{150}{\deg^2} and deep tier of \SI{210}{\deg^2}.


Combining current data with the DESI ``active trigger'' (dynamical tile ordering) SNe, the distance from \lcdm~improves from $3.8\sigma$ to $5.9\sigma$ if the current best-fit dark energy parameter values don't change, and the Figure of Merit improves from $108$ to $171$, or by 60\%. Without assuming an additional external low-$z$ SN sample, we reach $5.3\sigma$ and $154$ FoM.

In comparison, the current status quo (passive trigger, rigid tile ordering) would only reach $5.1\sigma$ with and $4.5\sigma$ without additional low-$z$ SNe, making the potential confirmation of dynamical dark energy highly dependent on a well-calibrated low-$z$ sample and the best-fit value.\footnote{This raises a technical point: The distance from \lcdm\ is calculated as $\sqrt{\Delta\chi^2}$, while the plotted Gaussian $5\sigma$ contour in 2D corresponds to a $\Delta\chi^2=28.74$, or $\sqrt{\Delta\chi^2}\approx5.36$, so \lcdm\ can't be claimed to be excluded in the 2D parameter plane if the radial distance barely exceeds $5\sigma$.} By simply going from passive to active triggers, we almost triple the total number of SNe, increasing the projected distance from \lcdm\ by nearly one full standard deviation.

We show our findings for the different scenarios in a plot of $w_0-w_a$ error ellipses, including $5\sigma$ contours, in \Cref{fig:ellipses_DESI_SN}, and total numbers in \Cref{fig:total_objects}.

\subsection{The importance to evaluate combined surveys}

We also show the error ellipses, broken down per source, in \Cref{fig:ellipses_by_source}. Clearly, the new SNe do not offer tight constraints on their own. However, the SN error ellipse is nearly orthogonal to the BAO+CMB ellipse. This is because low-redshift, nearby SNe are primarily sensitive to the current equation of state $w_0$, while BAO and CMB constrain an integrated combination of $w_0$ and $w_a$ over a much larger redshift range.

Evaluated on a standalone FoM of just 0.04, there would be no chance our proposed survey would ever be considered -- the current Union3 dataset alone has a FoM two orders of magnitude higher -- even though the combined FoM is ultimately improved by almost 50. We thus recommend survey design decisions be made with combined forecasts in mind: Doing so allowed us to identify SNe~Ia with $z<0.3$ as a fast path to test dynamical dark energy.

\subsection{Impact of systematics on cosmology}
\begin{figure}[]
    \centering
    \includegraphics[width=0.49\textwidth]{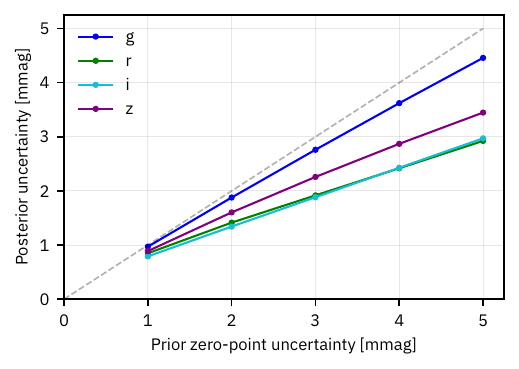}
    \caption{Zero-point calibration priors versus posteriors for each Rubin filter. The unified fitting framework allows for self-calibration, reducing posterior uncertainties by up to 40\% relative to the assumed priors. However, we find no floor beyond which the posterior uncertainty is always lower than the prior uncertainty for any realistic prior. This indicates a potential prior-dependence of cosmology results in hierarchical Bayesian frameworks even after self-calibration, and warrants further investigation.
    \label{fig:zp_priors}}
\end{figure}
\begin{figure*}[]
    \centering
    \includegraphics[width=0.99\textwidth]{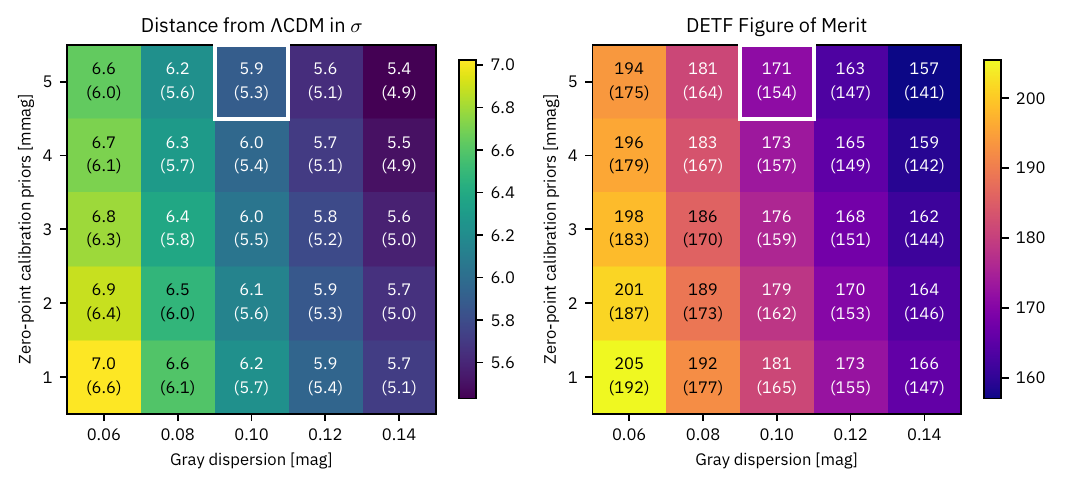}
    \caption{Distance from \lcdm\ in $\sigma$ (left) and Dark Energy Task Force Figure of Merit (right) as a function of the assumed zero-point calibration prior and gray dispersion — the two dominant systematics most likely to improve in the near future. Values in parentheses correspond to forecasts without a separate sample of new low-$z$ supernova. The $5\sigma$ threshold is reached or exceeded across a wide range of assumed priors. Neither systematic alone drives the final constraints.
    \label{fig:var_systematics}}
\end{figure*}

The sensitivity of frequentist dark energy analyses \citep{Brout2022Pantheon, DES5Y2024} to calibration and systematics has recently been pointed out by \citet{Efstathiou2025}. A re-analysis of the DES-5Y supernova sample \citep{DES5Y2024}, motivated by newer calibration methods \citep{Popovic2025a}, led to a change in the distance from \lcdm~by $1\sigma$ \citep{Popovic2025b}.

In a Bayesian framework such as \citet{Rubin2025Union3}, the dependence on the assumed calibration is partially absorbed by the model's improved ability to self-calibrate (\citet{Kim2006}, and described in \S\ref{sec:sims}). But even then, \citet{Rubin2025} shows that the Figure of Merit depends strongly on the assumed priors on zero-point calibration and fundamental color uncertainty, and weakly on wavelength uncertainty and the uncertainty on the count-rate nonlinearity (not applicable to Rubin CCD detectors).

Here, we check how cosmological constraints depend on the zero-point calibration and assumed dispersion: These are the most likely uncertainties to be improved. 

For example, recent filter re-calibration efforts for the Pantheon$+$ and DES-Y5 supernovae now achieve 2 to \SI{3}{mmag} internal precision per filter \citep{Popovic2025a}. LSST aims for internal calibration precision of \SI{1}{mmag} \citep{LSSTDESCSRD}. We allow the priors on the zero-point uncertainty to vary from 1 to \SI{5}{mmag}, with the latter as our currently-assumed baseline. Through the unified fitting process, the filters self-calibrate to varying degrees, leading to posterior zero-point uncertainties that are smaller by up to 40\%. This effect is broken down by filter in \Cref{fig:zp_priors}.

A potential ``point of stability'' towards which the posterior uncertainty choice would converge (after multiple iterations where the next priors are changed to the current posterior) does not seem to exist for any reasonable priors above \SI{1}{mmag}. Thus, the choice of calibration priors impacts the final cosmology, even when allowing for self-calibration. We also noticed that the assumed dispersion has little impact (sub-1\%) on the self-calibrated posterior uncertainty.

The remaining scatter in supernova magnitudes after standardization depends on the ability of the model to capture SN-to-SN variability, and is expected to decrease as models improve. We broke the typical scatter of \SI{0.15}{mag} down into a gray dispersion (\SI{0.10}{mag}) applied across the whole supernova, and per-filter color dispersion (\S\ref{sec:sims}) of \SI{0.03}{mag}. Here, we allow for the gray dispersion to vary broadly -- from an optimistic \SI{0.06}{mag}, currently only reachable with spectroscopic standardization methods (\S\ref{sec:outlook}), to \SI{0.14}{mag}, which would correspond to using a model trained on substantially noisier photometric data even when compared to current-generation measurements \citep{Rigault2025ZTFDR2, Rubin2025Union3}.

For these different scenarios, we compute the distances from \lcdm~in $\sigma$ and DETF FoMs, with (and without) the inclusion of a ZTF-like sample of low-$z$ supernovae. The results are summarized in \Cref{fig:var_systematics}. Distances from \lcdm~range from just below $5\sigma$ when excluding new low-$z$ SNe to $7\sigma$ in the optimal case. We find that there is not a single parameter driving the final uncertainties; rather, achieving further improved constraints depends on both (unexplained) dispersion and filter calibration.

\section{Potential Enhancements}\label{sec:outlook}

\subsection{Spectroscopic standardization \\ as an independent cross-check}

If a significant signal of dynamical dark energy were found, having a completely different SN standardization method arrive at the same conclusion would strengthen the discovery claim significantly. Spectroscopic standardization could offer a way to verify cosmology results independent of light curve analyses. 

\citet{Fakhouri2015} showed that ``twin'' SNe~Ia (SNe with similar spectra) have a similar absolute magnitude and that in general, subsets of SNe with more similar spectra have a reduced dispersion. Using this, ``twinning'' methods have been developed to standardize SNe~Ia using their near-peak spectra instead of their light curves. For example, Twins Embedding (\citealt{Boone2021a, Boone2021b}) generalizes the approach of \citet{Fakhouri2015} by projecting near-peak spectra onto a lower-dimensional embedding which describes intrinsic variabilities, and performs standardization based on that.  \citet{Stein2022} generalize this further, using the entire spectral time series to standardize with a Probabilistic Autoencoder (PAE) \citep{Boehm2020}.

Both Twins Embedding and PAE achieve significantly lower unexplained dispersions of \SI{0.07}{mag}. Even with noisier data, this has been shown to outperform traditional photometric standardization methods \citep{Rubin2025}.

These data-driven machine-learning-based approaches have several advantages -- standardization no longer relies on empirical astrophysical relations. Thus, astrophysical effects do not need to be considered separately. One example is the supernova mass step, a phenomenon describing the fact that SNe in massive galaxies appear systematically brighter \citep{Kelly2010, Sullivan2010} even after SALT standardization. With Twins Embedding, this systematic was not found post-standardization, suggesting that the data-driven model already accounts for the mass step. Recently, \citet{Ganot2025} demonstrated that even with low signal-to-noise spectra from ZTF, there was no mass step observed after standardization via Twins Embedding, even though with that observing program the overall spectral quality proved to be too low to improve the unexplained dispersion.

\texttt{SpecSim} (\S\ref{sec:DESI}) predicts that SN spectra taken with DESI will almost always significantly exceed the signal-to-noise values of the ZTF spectra used in \citet{Ganot2025}---a highly promising sign for the viability of twinning. However, our current SNR estimates do not yet account for the noise introduced by host galaxy subtraction for a separate later DESI observation. A discrete spectral subtraction of the host light (here DESI's very accurate fiber placement capability would prove critical, \citet{Guy2023DESI}), but template-based host subtraction might prove viable given the millions of galaxies DESI has already observed.
We leave a detailed study of using DESI spectra for precise standardization, including realistic host galaxy modeling, to future work.

\begin{figure*}[]
    \centering
    \begin{minipage}[t]{0.49\textwidth}
        \centering
        \includegraphics[width=\linewidth]{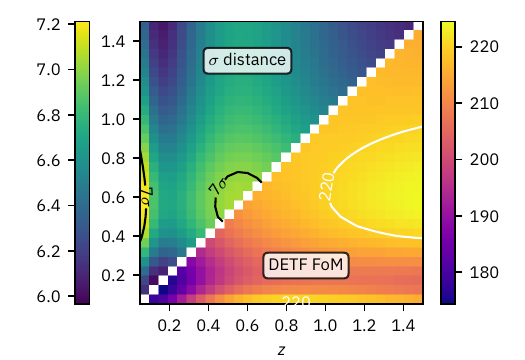}
        \caption{Distance from \lcdm~and Figure of Merit when adding supernovae at two redshift bins ($z_1$ and $z_2$) with $10\rm mmag$ total uncertainty each to baseline data (\S\ref{sec:toymodel}) plus DESI ``active trigger'' SNe. Compared to Figure \ref{fig:z_grid_current+lowz}, optimizing for FoM and distance from \lcdm~result in different optimal redshift combinations. However, both cases prefer to have a medium-redshift sample at $z\approx 0.6$.}
        \label{fig:2dgrid_DESI}
    \end{minipage}%
    \hfill
    \begin{minipage}[t]{0.49\textwidth}
        \centering
        \includegraphics[width=\linewidth]{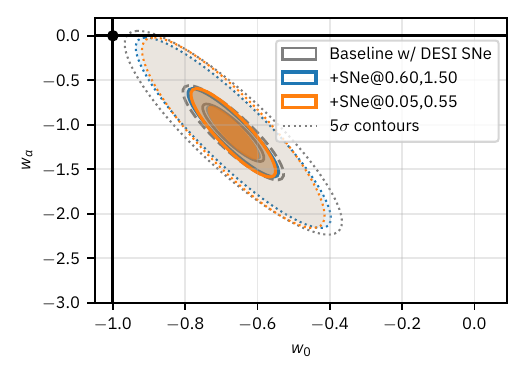}
        \caption{Projected $1, 2$ and $5\sigma$ contours after adding two hypothetical supernova measurements at the redshifts that optimize either the Figure of Merit (blue) or the $\sigma$-distance from \lcdm\ (orange), on top of the new baseline (\Cref{fig:ellipses_DESI_SN}). Optimizing for FoM shrinks the error ellipse but leaves it pointing toward \lcdm. Optimizing for $\sigma$-distance produces a larger error ellipse overall, but tilts contours further from \lcdm.}
        \label{fig:future_ellipses}
    \end{minipage}
\end{figure*}
\subsection{Further synergies and future surveys}

\subsubsection{Training Rubin pipelines}

The number of transients spectroscopically followed up by a one-year DESI active trigger program would exceed current photometric classification training datasets by an order of magnitude. Such a sample would serve as both a
stress test and a major expansion of the training data for Rubin's photometric redshift and classification pipelines. TiDES \citep{Frohmaier2025TiDES} will ultimately obtain more objects, but over five years and with selection effects; early improvements from a DESI program could propagate to all cosmology results over Rubin's projected ten-year lifetime.

\subsubsection{Optimal redshifts for $w_0$ and $w_a$ in the future}

If we find strong evidence against \lcdm, the next step would look beyond the simple $w_0-w_a$ parametrization of dynamical dark energy. However, here we complete the $w_0-w_a$ story by repeating the exercise of \S\ref{sec:toymodel}, but now asking which additional supernovae synergize best with the projected DESI-Rubin dataset (plus the baseline from \S\ref{sec:toymodel}). Such re-optimization for $\sigma$-distance from \lcdm\ and Figure of Merit points to the next avenue for improving dark energy constraints with SNe~Ia.
\Cref{fig:2dgrid_DESI} 
shows that the $\sigma$-distance is maximized by a combination of a very low-redshift sample at $z\approx 0.05$ and a mid-redshift sample at $z\approx 0.55$. The Figure of Merit is maximized by mid-redshift SNe at $z\approx 0.6$ and high-redshift SNe at $z\gtrsim 1.5$.

The resulting error ellipses of \Cref{fig:future_ellipses} illustrate why: optimizing for FoM shrinks the ellipse but does not change its orientation, so its semi-major axis still points toward \lcdm. Optimizing for $\sigma$-distance instead constrains $w_0$ more tightly, tilting the error ellipse away from \lcdm. This is expected: the low-$z$ component of the $\sigma$-optimal combination probes late-time dark energy, which is primarily encoded in $w_0$. 

Evaluating both metrics thus seems useful when designing future surveys; optimizing for both may yield the most robust dark energy measurements.\footnote{Purely optimizing for the $\sigma$-distance is speculative, as the best-fit point may shift with future data. But optimizing for FoM carries its own risk: one might end up with an error ellipse whose semi-major axis points directly at \lcdm, weakening a potential discovery claim.}

Targeting a combination of very low-, medium- ($z\approx 0.6$), and high-redshift supernovae thus appears most informative. High-redshift ($z>1$) SNe will be measured by space telescopes such as Roman \citep{Spergel2015Roman} and Lazuli \citep{Roy2026Lazuli}. Hence, we now discuss how the low- and medium-redshift tiers could be tackled with DESI and Rubin.

\subsubsection{Low redshifts: Rubin micro and twilight surveys}

Spectroscopy of very low-redshift supernovae ($z\lesssim 0.1$) are easily tractable from the ground. A small Rubin micro survey\footnote{\url{https://pstn-055.lsst.io/}; defined as $<3\%$ of survey time.} targeting the DESI-II fields north of the WFD border could cover \SI{500}{deg^2} per night in a single filter with short exposures in under ten minutes, corresponding to just 0.2\% of total survey time for a one-year campaign.\footnote{Rubin's \SI{9.6}{deg^2} FoV enables it to observe \SI{3450}{deg^2} per hour at 6 pointings per minute. After overhead, this leaves 3 to 4\,s exposures per pointing -- more than enough for SNe at $z<0.1$.} The resulting imaging data could also aid DESI-II target selection.

Alternatively, a Rubin twilight program (X. Tang et al., in prep) would follow nearby, bright supernovae during times that are outside the main Rubin survey window each night.

In either case, classification with DESI requires only $t_{\rm eff}<\SI{30}{s}$, so these objects could even be observed with what DESI refers to as a backup tile \citep{Schlafly2023}.

\subsubsection{Medium redshifts: Deep Drilling Field follow-ups}

Medium-redshift supernovae at $z\approx 0.5$--$0.6$ sit in a gap that a dedicated DESI follow-up program targeting Rubin's Deep Drilling Fields (DDFs) could help fill. Time dilation keeps supernovae near peak for longer in the observer frame, and we estimate with \texttt{SpecSim} (\S\ref{sec:DESI}) that two three-hour exposures per month suffice to continuously classify faint SNe~Ia at $z=0.5$, increasing to two five-hour exposures for faint SNe~Ia at $z=0.55$. Such a program would yield 100 spectroscopically-confirmed SNe~Ia per year, many at the most informative redshifts around $z=0.5$.

\subsubsection{DESI-II}

A potential DESI-II survey would cover a smaller footprint with good Rubin overlap over a projected six-year duration (\Cref{fig:footprint}). A supernova program analogous to the one proposed here would fit naturally into the first two years, when most tiles are still available and SNe can actively trigger observations with few restrictions --- consistently targeting tiles with multiple SNe per pointing. Despite the smaller footprint, we estimate that DESI-II could follow up 7000 SNe~Ia near peak per year until tile depletion becomes a limiting factor. DESI-II could also continue a DDF follow-up program.
\section{Conclusions} \label{sec:Conclusion}

We investigated a fast path to test dynamical dark energy versus \lcdm\ at the $>5\sigma$ level. Considering current CMB, BAO and SN data, we found that Type Ia supernovae provide good leverage on $w_0$ and $w_a$ at redshifts near dark energy-matter equality (\S\ref{sec:toymodel}). We considered two major Stage-IV cosmology experiments -- the Dark Energy Spectroscopic Instrument and the Vera C. Rubin Observatory -- that share the common goal of measuring dynamical dark energy, and found that they synergize well in this redshift range.

We designed a DESI spectroscopic follow-up program of Rubin-discovered supernovae at essentially no added observing cost: The total number of tile visits is fixed by DESI's main survey; supernovae simply act as triggers that reorder the queue, allowing SN spectra to be measured opportunistically, all while the main survey completes as planned. The overlap of DESI and Rubin's Wide Fast Deep rolling cadence in 2027 provides a window of at least one year to execute this program (\S\ref{sec:DESI}).

Under baseline assumptions (\S\ref{sec:toymodel}, \S\ref{sec:sims}), our forecast reaches a $5.9\sigma$ distance from \lcdm~at the current best-fit values for $w_0$ and $w_a$ (\S\ref{sec:Results}). We obtained these results with a unified fitter that takes into account the most important systematics and their correlations.

A $5\sigma$ result is necessary but not sufficient for a credible discovery claim. Few ``moving parts'' as well as independent cross-checks are important for credibility. Our program provides both: it eliminates common systematics from selection effects, photometric redshift uncertainties and failures, and photometric classification biases. Additionally, the possibility for spectroscopic standardization not only offers a cross-check of results from traditional analyses that use empirical light-curve fitters, but could even improve precision through reduced per-supernova scatter.

A methodological lesson emerges from this work. The standalone Figure of Merit of our proposed survey is two orders of magnitude smaller than existing supernova compilations, yet it yields a 50\% improvement when combined with current data. Evaluating surveys and probes in isolation can thus mislead when it comes to survey design and prioritization, and we therefore recommend holistic assessments that include all existing datasets.

Beyond dynamical dark energy measurements, this sample will provide training data for photometric classifiers and photo-$z$ estimators, and thus will serve as a timely stress test of Rubin pipelines. Early improvements prompted by this cross-check could compound over Rubin's full ten-year lifetime and influence a whole generation of cosmology measurements.

DESI follow-up of Rubin supernovae from a combination of the Wide Fast Deep (this work), a potential low-$z$ program, and the Deep Drilling Fields would provide a unique opportunity (\S\ref{sec:outlook}): Supernova cosmology could be performed with a clean, spectroscopically confirmed, homogeneous sample, without results hinging on the difficult task of cross-calibrating across different photometric systems. We intend to study such a comprehensive supernova program in future work. 

\begin{acknowledgments}
We thank John Banovetz and Samuel Brieden for their detailed and helpful comments on the manuscript and insights into DESI operations, and Xander Hall, Antonella Palmese, Eric Linder and Julien Guy for useful discussions. This work was supported in part by the Director, Office of Science, Office of High Energy Physics of the US Department of Energy under contract No. DE-AC025CH11231.
\end{acknowledgments}

\begin{contribution}
JT led the detailed analysis, ran the simulations, and wrote the bulk of the paper. GA conceived the use of DESI in the active trigger mode, developed initial feasibility calculations, and provided regular and detailed input on the analysis and manuscript. SP supervised the project and provided regular feedback on the analysis and manuscript. DR assisted with running the distance modulus covariance matrix code and provided feedback on the analysis and manuscript. DS provided extensive insights into DESI at all stages of this project and provided feedback on the analysis and manuscript.

\end{contribution}

%

\software{\texttt{astropy} \citep{astropy:2013, astropy:2018, astropy:2022}, \texttt{numpy} \citep{numpy}, matplotlib \citep{matplotlib}, 
\texttt{SNCosmo} \citep{Barbary2016}, \texttt{wfirst-sim} \citep{Rubin2025}.
          }



\bibliography{bib}{}
\bibliographystyle{aasjournalv7}

\appendix
\section{Fisher matrix formalism}
\label{sec:fisher}

We use the Fisher information matrix \citep{Fisher1922} to estimate the Gaussian covariance of cosmological parameters $\bm p$. It depends on the assumed Gaussian measurement uncertainties, given by a covariance matrix $\mathcal{C}_{ij}$ of the observations $\bm O$, and the choice of the fiducial cosmology.

The Fisher matrix is
\begin{align}
    F_{ab} = \sum_{i,j} \frac{\partial O_i}{\partial p_a}\Bigr|_{\rm fid}\,\mathcal{C}_{ij}^{-1}\,\frac{\partial O_j}{\partial p_b}\Bigr|_{\rm fid}
\end{align}
where the partial derivatives are evaluated at some chosen fiducial cosmology.

$O_i$ refers to any individual measurement of any observable (e.g. supernova magnitude, distance to last scattering, etc.). Practically, measurements of different cosmological observables (such as from the CMB, BAO, SNe) are uncorrelated, so their Fisher matrices can simply be combined additively. 

The covariance of cosmological parameters from combining different cosmological probes is then the inverse of the combined Fisher matrix of these probes.

For supernovae, the observable is the distance modulus 
\begin{align}
    \mu = 5\log_{10}\mathcal{D}_L(z|w_0,w_a,\Omega_M)+\mathcal{M}
\end{align}
where $\mathcal{D}_L=H_0d_L/c=(1+z)\int_0^z\frac{dz'}{E(z')}$ is the dimensionless luminosity distance; and the dimensionless Hubble parameter for a flat universe is
\begin{equation}
    E(z)=\frac{H(z)}{H_0}=\sqrt{\Omega_m (1+z)^3 + (1-\Omega_m) (1+z)^{3(1+w_0+w_a)} e^{\frac{-3w_a z}{1+z}}}.
\label{eq:E}
\end{equation}
For dark energy cosmology, $\mathcal{M}$ is commonly treated as a nuisance parameter, so dark energy measurements do not explicitly depend on $H_0$. $\mathcal{M}$ is typically treated as a nuisance parameter \citep{Perlmutter1997}.

The corresponding 4$\times$4 Fisher matrix, where the rows and columns correspond to the information on the parameters $\bm{p}=\{w_0,w_a,\Omega_M,\mathcal{M}\}$, is then given by
\begin{align}
    F_{ab}^{\rm SN} = \sum_{i,j} \frac{\partial \mu(z_i)}{\partial p_a}\Bigr|_{\rm fid}\,\mathcal{C}^{-1}(z_i,z_j)\,\frac{\partial \mu(z_j)}{\partial p_b}\Bigr|_{\rm fid}
    \label{eq:FisherSN}
\end{align}
where $\mathcal{C}$ is now the distance modulus covariance matrix with units $\rm mag^2$.




\subsection{Combining new SNe with existing data}

When combining with measurements of Baryon Acoustic Oscillations (BAO) and the Cosmic Microwave Background (CMB), we use the parameter covariance matrices released with DESI DR2 \citep{DESIDR2BAO2025}.\footnote{\url{https://data.desi.lbl.gov/public/papers/y3/bao-cosmo-params/cobaya/base_w_wa/}}

When combining new SNe with BAO+CMB+Union3, we have to be a bit careful. Even though the parameter covariance matrix was obtained using Union3 data, the released covariance matrix does not contain information about $\mathcal{M}$ explicitly because it is marginalized over.\footnote{Since $\mathcal{M}$ appears as a linear term in $\mu$, it can be analytically marginalized over in the likelihood. Typically, a Gaussian with zero mean and $\sigma_\mathcal{M}\rightarrow \infty$ is assumed in the process. However, the lack of $\mathcal{M}$ information leads to practical problems when adding supernovae in a very limited redshift range -- no matter how precise these measurements are, they could not improve constraints on $w_0-w_a$ in this scenario.} So when pre-processing like above, $\mathcal{M}$ is under-constrained. To alleviate this, we compute the Union3 Fisher matrix and insert its $\mathcal{M}$-information into the Fisher matrix.

This somewhat artificial modification of the Fisher matrix has little impact on all relevant metrics. When inverting this Fisher matrix back to get a (now slightly modified) parameter covariance matrix, changes are small: $<0.5\%$ in the Figure of Merit and distance from \lcdm, and $<1\%$ for all individual entries in the $w_0,w_a,\Omega_M$ sub-covariance matrix.

\subsection{Figure of Merit and distance from \lcdm} \label{sec:FoM}

One obtains the covariance matrix of cosmological parameters by inverting the combined Fisher matrix. The Figure of Merit as defined by the Dark Energy Task Force (DETF FoM, \citealt{Albrecht2006DETF}) is related to the inverse area of the $w_0-w_a$ ellipse,
\begin{align}
    \label{eq:FoM}
    \mathrm{FoM} = \frac{1}{\sqrt{\det \mathrm{Cov}(w_0, w_a)}},
\end{align}
and is commonly used as a measure of $w_0-w_a$ constraints: The smaller the ellipse, the larger the FoM is, and the better the constraints.

While the covariance matrix of the observations is in principle cosmology-independent, the Fisher matrix (and thus the parameter covariance matrix) depends on the choice of the fiducial cosmology (see eq. \ref{eq:FisherSN}). Typically, the same experiment will lead to tighter constraints (higher FoM) when assuming \lcdm~as one's fiducial cosmology.

We compute the distance in $\sigma$ from \lcdm~($w_0 = -1$, $w_a = 0$) via the Mahalanobis distance
\begin{equation}
    \sigma = \sqrt{
    \left( \Delta w_0 \;\; \Delta w_a \right)
    \,\mathrm{Cov}(w_0, w_a)^{-1}\,
    \begin{pmatrix}
        \Delta w_0 \\[2pt] \Delta w_a
    \end{pmatrix}
    },
\end{equation}
where $\Delta w_0$ and $\Delta w_a$ are the deviations of the fiducial point from \lcdm.

\section{Simulations and systematics} 
\label{sec:Simulations}

\subsection{Supernova generation and light curve simulation}

Inside the duration and footprint of the simulated survey, redshifts and times at maximum light of supernovae are drawn based on volumetric rates in \citet{Rodney2014}.

Supernova redshifts are binned in bins of $0.05$ width for $z>0.05$. The lowest redshift bin center is at $z=0.075$. SNe below $z=0.05$ are not generated at all. This is inherited from \citet{Rubin2025}, who made this choice since peculiar velocities due to bulk flows in the cosmic web have an especially significant impact on the measured distance moduli below this redshift, and peculiar velocities are not accounted for in the simulation. So including these SNe would lead to overly optimistic results.

For each supernova, SALT3 \citep{Kenworthy2021, Pierel2022} light-curve parameters are drawn, and \texttt{SNCosmo} \citep{Barbary2016} is used to generate supernova SEDs (spectral-energy distributions) based on SALT3-NIR \citep{Pierel2022} from $-15$ to $+45$ days from maximum in the rest frame of the supernova. This information is then used to generate a pixelized image.

Given the input cadences and exposure times for each filter, measurements of fluxes and uncertainties of the light curve are generated from these simulated images. Figure \ref{fig:median_LC} shows an example light curve.

\subsection{Model fitting error}

An important contribution to the total uncertainty comes from the supernova model. The model fits the measured light curves to standardize the supernovae. The quality of the fits can be influenced by the signal-to-noise ratios of individual measurements, but also uncertainties in the calibration of the instruments. Suboptimally calibrated models produce biased distance modulus predictions, which can then systematically affect the final cosmology results. The systematic uncertainty introduced by imperfect model fitting is a significant component in the total error budget \citep{Rubin2025Union3}.

\subsection{Fitting process and calculation of distance modulus covariance matrix}

We use a unified Bayesian fitting framework similar to UNITY1.5 \citep{Rubin2025Union3}. We fit all cosmological parameters and nuisance parameters (like the previously mentioned calibration uncertainties) simultaneously. By solving for all parameters at the same time, we propagate the uncertainties from all sources into our final cosmology result.

Our approach is essentially to minimize a global $\chi^2$
\begin{align}
    \chi^2 = \sum_{i=1}^{N_{\text{SN}}} \chi^2_{i, \text{data}} + \chi^2_{\text{priors}}
\end{align}
with a per-supernova
\begin{align}
    \chi^2_{i, \text{data}} = \left( \bm{\mu}_{\text{obs}, i} - \bm{\mu}_{\text{model}, i} \right)^\top \bm{C}_i^{-1} \left( \bm{\mu}_{\text{obs}, i} - \bm{\mu}_{\text{model}, i} \right)
\end{align}
where the $\bm{\mu}_{\text{obs}, i}$ vector contains all individual light-curve point measurements of the $i$-th supernova, and $\bm{C}_i$ their statistical weights. This covariance depends on the signal-to-noise ratio of an individual measurement (on the diagonal), and the per-SN-effects (intrinsic dispersion and dust extinction uncertainty). $\bm{\mu}_{\text{model}, i}$ depends on the theoretical prediction of the distance modulus $\mu_{\text{th}, i}$ for that supernova, and the calibration, including the supernova model. In contrast to frequentist analyses, these are not assumed rigid and uncorrelated, but are also found during the fit.\footnote{For a detailed breakdown of $\bm{\mu}_{\text{model}, i}$, we refer the reader to \citet{Rubin2025}, where $\bm{\mu}_{\text{model}, i}$ is called $m_{ijk}$.}

We include Gaussian priors with zero mean on the instrumental calibration in $\chi^2_{\text{priors}}$ to prevent the fit from assuming unreasonable calibration values.

$\chi^2$ is a function of all distance moduli and nuisance parameters. We find the global minimum and compute the Hessian at that point. By inversion, we obtain a full covariance matrix of distance moduli and nuisance parameters. The distance modulus covariance matrix $\mathcal{C}(z_i,z_j)$ (cf. eq. \ref{eq:FisherSN}) is then extracted as a sub-matrix of the full covariance matrix. This process marginalizes over all nuisance parameters and thus correctly propagates all uncertainties into $\mathcal{C}(z_i,z_j)$.

The supernova model is parametrized and integrated into $\bm{\mu}_{\text{model}, i}$. There is a flat prior placed on these parameters. The curvature of $\chi^2$ as a function of model fitting parameters is automatically found during the fit, so the model error is included in the full covariance matrix. Thus, it is propagated to the distance modulus covariance matrix as intended.

\subsubsection{Cosmology forecast}

Because $\chi^2$ is a function of all distance moduli and calibration/model parameters, the distance moduli are correlated. We find the global minimum of $\chi^2$ and compute the Hessian at that point. By inversion, we obtain a full covariance matrix of distance moduli and nuisance parameters. The distance modulus covariance matrix $\mathcal{C}(z_i,z_j)$ is then extracted as a sub-matrix of the full covariance matrix. This process marginalizes over all nuisance parameters and thus correctly propagates all uncertainties into $\mathcal{C}(z_i,z_j)$.

The Fisher matrix of the simulated survey is calculated based on the simulated distance modulus covariance matrix $\mathcal{C}(z_i,z_j)$ (\cref{eq:FisherSN}). After combining this with the baseline Fisher matrix (\S\ref{sec:toymodel}), we invert the combined Fisher matrix to obtain the covariance matrix of cosmological parameters.

\section{Toy model to calculate supernova Ia yield}
\label{sec:SNrates}

We construct a simple toy model to estimate the expected ratio of total observed supernovae to Type Ia supernovae, $N_{\rm tot}/N_{\rm Ia}$, in a magnitude-limited survey. The number of supernovae of a given type, $N$, that a survey can detect is proportional to the intrinsic volumetric rate, $r$, and the maximum volume, $V_{\rm max}$, out to which detection is possible. This volume is determined by the survey's limiting magnitude, $m_{\rm lim}$, and the supernova's peak absolute magnitude, $M_{\rm peak}$. From the definition of the distance modulus, the maximum detectable distance $d_{\rm max} \propto 10^{0.2(m_{\rm lim} - M_{\rm peak})}$, which means the corresponding volume scales as $V_{\rm max} \propto d_{\rm max}^3 \propto 10^{0.6(m_{\rm lim} - M_{\rm peak})}$ in an Euclidean geometry approximation.

When we take the ratio of the number of observed Core-Collapse Supernovae ($N_{\rm CC}$) to Type Ia Supernovae ($N_{\rm Ia}$), the survey-dependent terms such as the limiting magnitude cancel out. This leaves a ratio that depends only on the intrinsic rates and luminosities of the two populations:
\begin{align}
    \frac{N_{\rm CC}}{N_{\rm Ia}} = \frac{r_{\rm CC} \times V_{\rm max, CC}}{r_{\rm Ia} \times V_{\rm max, Ia}} = \frac{r_{\rm CC}}{r_{\rm Ia}} \times 10^{\left[ 0.6(M_{\rm Ia}-M_{\rm CC}) \right]}
\end{align}
To evaluate this ratio, we adopt representative values from the literature. At a redshift of $z \approx 0.3$, the supernova rates are approximately $r_{\rm Ia} \approx \SI{3e-5}{yr^{-1}.Mpc^{-3}}$ \citep{Frohmaier2019} and $r_{\rm CC} \approx \SI{2e-4}{yr^{-1}.Mpc^{-3}}$ \citep{Strolger2015}. For the peak absolute magnitudes, we use a standard value of $M_{\rm Ia} = -19.3$ for SNe Ia. For the fainter and more diverse CCSN population, we adopt a characteristic absolute magnitude of $M_{\rm CC} = -17.0$, a value consistent with the mean of observed distributions \citep{Richardson2014}.

Substituting these values into the equation yields:
\begin{align}
    \frac{N_{\rm CC}}{N_{\rm Ia}} & \approx \frac{\SI{2e-4}{}}{\SI{3e-5}{}} \times 10^{\left[ 0.6(-19.3 - (-17.0)) \right]} \nonumber \\
    & \approx 6.67 \times 10^{\left[ 0.6(-2.3) \right]} \nonumber \\
    & \approx 6.67 \times 10^{-1.38} \approx 0.28
\end{align}
This result suggests that the number of observable CCSNe is approximately 28\% of the number of SNe Ia. Therefore, the ratio of the total number of supernovae to SNe Ia is $N_{\rm tot}/N_{\rm Ia} = (N_{\rm Ia} + N_{\rm CC})/N_{\rm Ia} \approx 1.28$.

\section{Exposure time calculations and custom scheduling}
\label{sec:customscheduling}

Here we present a few preliminary calculations and concepts revolving around a custom DESI supernova program, using custom exposure times and without being restricted to tiles. This would become applicable for a possible extension to the baseline survey presented in the main body of this work.

In short, one would run a clustering algorithm to find pointings with the most SNe, and then calculate the required exposure time to reach the faintest object in that pointing.

\subsection{Impact of SN brightness distribution}

Intrinsically fainter supernovae are exponentially harder to observe, so determining the exposure time required for an average supernova is not representative of the actually required average exposure time. Here, we calculate the average additional time required to observe a supernova, allowing for a distribution in raw observed brightnesses (not to be confused with the intrinsic scatter after standardization).

We first determine the additional exposure time required to maintain a constant signal-to-noise ratio for a fainter supernova. The signal scales like ($S \propto F \, t$). We observe from the ground, where the sky background will dominate, especially for fainter supernovae for which this calculation is relevant. Here, the noise scales like ($N \propto \sqrt{t}$).

To maintain a constant SNR, we thus require $F\sqrt{t} = F'\sqrt{t'}$, or
\begin{equation}
    \frac{t'}{t} = \left(\frac{F}{F'}\right)^2 = 10^{0.8 \Delta m}
\end{equation}
where we used $\Delta m=-2.5\log_{10}(F'/F)$.

Fainter supernovae thus require exponentially more exposure time. Thus, even with a symmetrical distribution of magnitudes, the average required exposure time $\left\langle t\right\rangle$ is skewed upwards compared to the time $t_0$ required to expose a supernova with average magnitude. Assuming $\Delta m$ follows a Gaussian distribution $\mathcal{N}(0,\sigma)$, the expectation value is
\begin{equation}
     \frac{\left\langle t\right\rangle}{t_0}  = \left\langle 10^{0.8 \Delta m} \right\rangle = \exp\left[ \frac{1}{2} \sigma^2 (0.8 \ln 10)^2 \right].
\end{equation}

For the simplest standardization model only taking into account color $c$ and stretch $x_1$, the corrected magnitude is
\begin{align}
    m_{\rm corr} = m_B + \alpha x_1 - \beta c+...
\end{align}
where $m_B$ is the observed magnitude in the $B$-band \citep{Tripp1998}.

Typical values are $\sigma_{x_1}=1$ and $\alpha\approx 0.17$, $\sigma_c\approx 0.09$ and $\beta\approx 3.6$ \citep{Rubin2025Union3} and a residual scatter/magnitude dispersion of $\sigma_{m,{\rm corr}}=0.15$. ``Error propagating'' these to $m_B$ leads to a scatter of raw observed magnitudes $\sigma_{m}\approx0.40$. Thus, $\left\langle t\right\rangle/t_0 \approx1.30$, so the average required exposure time is around $30\%$ higher than the exposure time needed for an average-brightness supernova.



Here we use a middle-ground assumption of $\sigma_{m}\approx0.35$ raw scatter in our exposure time analysis, resulting in $\left\langle t\right\rangle/t_0\approx1.23$.

Since this effect is not explicitly redshift dependent (if the SN population doesn't drift), $\left\langle t\right\rangle/t_0$ can just be applied as a global modifier to the total exposure time.

\subsection{Impact of redshift binning}

Assuming all supernovae to be at the bin center will lead to an underestimate of exposure time: Just by volumetric arguments, there will be more supernovae per unit time in the redshift interval $[z_{\rm mid}, z_{\rm mid}+\Delta z/2]$ compared to $[z_{\rm mid}-\Delta z/2, z_{\rm mid}]$.

Given an approximately constant volumetric rate, the number of supernovae per unit time $dN$ in a ``comoving distance shell'' of thickness $d\chi$ is
\begin{align}
    dN &\propto \chi^2(z) \, d\chi \, \frac{1}{1+z}
\end{align}
where again $\chi(z)=\int_0^z\frac{dz'}{E(z')}$ and $E(z)$ given in eq. \ref{eq:E}. The additional redshift factor accounts for the time dilation that converts source-frame rate to observer-frame rate.

The quantity $dN/dz$ can be interpreted as an unnormalized probability density function to find a supernova at some redshift $z$. From the above equation, using $d\chi/dz=1/E(z)$, and omitting constants, we get
\begin{align}
    p(z) \propto \frac{\chi^2(z)}{(1+z)E(z)}.
\end{align}

Clearly, more supernovae are found at higher redshifts, so the mean redshift of a supernova in any given bin is higher than its center. To compute the impact on the required exposure time, we again use the ansatz that the signal-to-noise ratio should stay constant. This approach only works here because the redshift bins are narrow, so we don't lose significant parts of the the spectrum from redshifting out of the range of the filters.\footnote{For bins of width $\Delta z=0.05$, the spectra are redshifted by less than $5\%$, and this difference gets smaller towards higher redshifts.}

The ratio of the average exposure time in a redshift bin $\left\langle t\right\rangle$ to exposure time at the redshift bin center $t_{\rm mid}$ is
\begin{equation}
    \frac{\left\langle t\right\rangle_z}{t_{\rm mid}} = \frac{\left\langle d_L^4(z)\right\rangle_z}{d_L^4(z_{\rm mid})}
\end{equation}
where the expectation value integral is performed from $z_{\rm{min}}$ to $z_{\rm{max}}$, the borders of each redshift bin. This effect is smaller at higher redshifts due to the smaller relative bin width $\Delta z/z$.

\begin{table}[]
    \centering
    \begin{tabular}{c|c cc |c}
    $z$-bin & $t_0\,$[min] & $\rm SNR_{15\si{\angstrom}}$ & $\rm SNR_{tot}$ & $\langle t \rangle\,$[min] \\
    \hline
    $[0.05, 0.10]$ & 0.5 & 5.8-6.6 & 106-119 & 1.0 \\
    $[0.10, 0.15]$ & 1.5 & 5.9-6.8 & 105-118 & 2.2 \\
    $[0.15, 0.20]$ & 3.0 & 5.6-6.4 & 96-109 & 4.1 \\
    $[0.20, 0.25]$ & 5.5 & 5.2-6.1 & 93-105 & 7.2 \\
    $[0.25, 0.30]$ & 9.5 & 5.0-5.9 & 90-103 & 12.2 \\
    $[0.30, 0.35]$ & 14.5 & 5.0-5.5 & 88-96 & 18.4 \\
    \end{tabular}
    \caption{Required effective exposure times $t_0$ to observe a SN Ia with average brightness at the redshift bin center with $\rm SNR_{15\si{\angstrom}}>5$. SNR ranges refer to $\pm 5$ days (observer frame) around peak brightness. $\langle t \rangle_z$ is the mean exposure time accounting for the redshift distribution and magnitude scatter within the bin.}
    \label{tab:exposure_times}
\end{table}

\end{document}